\documentclass[aps,prd,floats,floatfix,nofootinbib]{revtex4-1}
\usepackage{graphicx}
\usepackage{amssymb}
\usepackage{epstopdf}
\usepackage{hyperref}

\usepackage{amsmath}

\begin{document}

\title{Characterizing a benchmark scenario for heavy Higgs boson searches in the Georgi-Machacek model}

\author{Heather E.~Logan}
\email{logan@physics.carleton.ca} 

\author{Mark B.~Reimer}

\affiliation{Ottawa-Carleton Institute for Physics, Carleton University, 1125 Colonel By Drive, Ottawa, Ontario K1S 5B6, Canada}

\date{September 6, 2017}                                 

\begin{abstract}
The Georgi-Machacek model is used to motivate and interpret LHC searches for doubly- and singly-charged Higgs bosons decaying into vector boson pairs.  In this paper we study the constraints on and phenomenology of the ``H5plane'' benchmark scenario in the Georgi-Machacek model, which has been proposed for use in these searches.  We show that the entire H5plane benchmark is compatible with the LHC measurements of the 125~GeV Higgs boson couplings.  We also point out that, over much of the H5plane benchmark, the lineshapes of the two CP-even neutral heavy Higgs bosons $H$ and $H_5^0$ will overlap and interfere when produced in vector boson fusion with decays to $W^+W^-$ or $ZZ$.  Finally we compute the decay branching ratios of the additional heavy Higgs bosons within the H5plane benchmark to facilitate the development of search strategies for these additional particles.
\end{abstract}

\maketitle

\section{Introduction}

Since the discovery of a Standard Model (SM)-like Higgs boson at the CERN Large Hadron Collider (LHC) in 2012~\cite{Aad:2012tfa}, much experimental and theoretical attention has been devoted to testing the possibility that the Higgs sector contains additional scalars beyond the single SM isospin doublet.  An interesting possibility among these extensions is that part of electroweak symmetry breaking---and hence part of the masses of the $W$ and $Z$ bosons---could be generated by scalars in isospin representations larger than the doublet.  A prototype model in this class is the Georgi-Machacek (GM) model~\cite{Georgi:1985nv,Chanowitz:1985ug}, which contains a real and a complex isospin-triplet scalar in addition to the usual SM Higgs doublet.  

A key feature of the GM model is the presence of doubly- and singly-charged Higgs bosons, $H_5^{\pm\pm}$ and $H_5^{\pm}$, that couple to SM vector boson pairs with an interaction strength proportional to the vacuum expectation value (vev) of the triplets.  Constraining this coupling therefore directly constrains the allowed contribution of the triplets to the masses of the $W$ and $Z$ bosons.  LHC searches for these scalars have been performed with production via vector boson fusion and decays to a pair of vector bosons~\cite{Khachatryan:2014sta,Aad:2015nfa,Sirunyan:2017sbn}; the LHC measurement of the like-sign $W$ boson cross section in vector boson fusion~\cite{Aad:2014zda} also provides sensitivity to the doubly-charged scalar~\cite{Chiang:2014bia}.  When the branching ratios of $H_5^{\pm\pm}$ and $H_5^{\pm}$ to vector boson pairs are essentially 100\%, these searches directly constrain the triplet vev $v_{\chi}$ as a function of the common mass $m_5$ of these scalars.

To aid the interpretation of these and future similar searches, the LHC Higgs Cross Section Working Group recently developed the ``H5plane'' benchmark scenario for the GM model~\cite{deFlorian:2016spz}.  The H5plane benchmark depends on two free input parameters, $m_5$ and $s_H \equiv \sqrt{8} v_{\chi}/v$ (where $v = (\sqrt{2} G_F)^{-1/2}$ is the SM Higgs vev), and the production cross sections for $H_5^{\pm\pm}$ and $H_5^{\pm}$ in vector boson fusion are proportional to $s_H^2$.  The other parameters of the model are fixed in the benchmark so that BR($H_5 \to VV) = 1$ to a very good approximation.  Predictions for the production cross sections (at next-to-next-to-leading order in QCD) and decay widths of these scalars have been provided in the context of the H5plane benchmark for LHC collisions at 8~\cite{Zaro:2015ika} and 13~TeV~\cite{deFlorian:2016spz}.

In this paper we perform the first comprehensive survey of the phenomenology of the H5plane benchmark in the GM model.  We show that the entire H5plane benchmark is compatible with the LHC measurements of the 125~GeV Higgs boson couplings from 7 and 8~TeV data~\cite{Aad:2016}.  We point out that, over much of the H5plane benchmark, the lineshapes of the two CP-even neutral heavy Higgs bosons $H$ and $H_5^0$ will overlap and interfere when these scalars are produced in vector boson fusion with decays to $W^+W^-$ or $ZZ$.  We also display the decay branching ratios of the additional heavy Higgs bosons within the H5plane benchmark to facilitate the development of search strategies for these additional particles.  Our numerical work is done using the public code {\tt GMCALC 1.2.1}~\cite{Hartling:2014xma}.

This paper is organized as follows.  In the next section we review the GM model and the specification of the H5plane benchmark.  Section~\ref{sec:properties} contains the bulk of our results.  We conclude in Sec.~\ref{sec:conclusions}.

\section{Georgi-Machacek model}
\label{sec:model}

The scalar sector of the GM model~\cite{Georgi:1985nv,Chanowitz:1985ug} consists of the usual complex doublet $(\phi^+,\phi^0)^T$ with hypercharge\footnote{We use $Q = T^3 + Y/2$.} $Y = 1$, a real triplet $(\xi^+,\xi^0, -\xi^{+*})^T$ with $Y = 0$, and  a complex triplet $(\chi^{++},\chi^+,\chi^0)^T$ with $Y=2$.  The doublet is responsible for the fermion masses as in the SM.
Custodial symmetry, required in order to avoid stringent constraints from the $\rho$ parameter, is preserved at tree level by imposing a global SU(2)$_L \times$SU(2)$_R$ symmetry on the scalar potential.
To make this symmetry explicit, we write the doublet in the form of a bidoublet $\Phi$ and combine the triplets into a bitriplet $X$:
\begin{equation}
	\Phi = \left( \begin{array}{cc}
	\phi^{0*} &\phi^+  \\
	-\phi^{+*} & \phi^0  \end{array} \right), \qquad
	X =
	\left(
	\begin{array}{ccc}
	\chi^{0*} & \xi^+ & \chi^{++} \\
	 -\chi^{+*} & \xi^{0} & \chi^+ \\
	 \chi^{++*} & -\xi^{+*} & \chi^0  
	\end{array}
	\right).
	\label{eq:PX}
\end{equation}
The vevs are given by $\langle \Phi  \rangle = \frac{ v_{\phi}}{\sqrt{2}} I_{2\times2}$  and $\langle X \rangle = v_{\chi} I_{3 \times 3}$, where $I_{n \times n}$ is the $n\times n$ unit matrix and the $W$ and $Z$ boson masses constrain
\begin{equation}
	v_{\phi}^2 + 8 v_{\chi}^2 \equiv v^2 = \frac{1}{\sqrt{2} G_F} \approx (246~{\rm GeV})^2.
	\label{eq:vevrelation}
\end{equation} 

The most general gauge-invariant scalar potential involving these fields that conserves custodial SU(2) is given, in the conventions of Ref.~\cite{Hartling:2014zca}, by\footnote{A translation table to other parameterizations in the literature has been given in the appendix of Ref.~\cite{Hartling:2014zca}.}
\begin{eqnarray}
	V(\Phi,X) &= & \frac{\mu_2^2}{2}  \text{Tr}(\Phi^\dagger \Phi) 
	+  \frac{\mu_3^2}{2}  \text{Tr}(X^\dagger X)  
	+ \lambda_1 [\text{Tr}(\Phi^\dagger \Phi)]^2  
	+ \lambda_2 \text{Tr}(\Phi^\dagger \Phi) \text{Tr}(X^\dagger X)   \nonumber \\
          & & + \lambda_3 \text{Tr}(X^\dagger X X^\dagger X)  
          + \lambda_4 [\text{Tr}(X^\dagger X)]^2 
           - \lambda_5 \text{Tr}( \Phi^\dagger \tau^a \Phi \tau^b) \text{Tr}( X^\dagger t^a X t^b) 
           \nonumber \\
           & & - M_1 \text{Tr}(\Phi^\dagger \tau^a \Phi \tau^b)(U X U^\dagger)_{ab}  
           -  M_2 \text{Tr}(X^\dagger t^a X t^b)(U X U^\dagger)_{ab}.
           \label{eq:potential}
\end{eqnarray} 
Here the SU(2) generators for the doublet representation are $\tau^a = \sigma^a/2$ with $\sigma^a$ being the Pauli matrices, the generators for the triplet representation are
\begin{equation}
	t^1= \frac{1}{\sqrt{2}} \left( \begin{array}{ccc}
	 0 & 1  & 0  \\
	  1 & 0  & 1  \\
	  0 & 1  & 0 \end{array} \right), \qquad  
	  t^2= \frac{1}{\sqrt{2}} \left( \begin{array}{ccc}
	 0 & -i  & 0  \\
	  i & 0  & -i  \\
	  0 & i  & 0 \end{array} \right), \qquad 
	t^3= \left( \begin{array}{ccc}
	 1 & 0  & 0  \\
	  0 & 0  & 0  \\
	  0 & 0 & -1 \end{array} \right),
\end{equation}
and the matrix $U$, which rotates $X$ into the Cartesian basis, is given by~\cite{Aoki:2007ah}
\begin{equation}
	 U = \left( \begin{array}{ccc}
	- \frac{1}{\sqrt{2}} & 0 &  \frac{1}{\sqrt{2}} \\
	 - \frac{i}{\sqrt{2}} & 0  &   - \frac{i}{\sqrt{2}} \\
	   0 & 1 & 0 \end{array} \right).
	 \label{eq:U}
\end{equation}

The physical fields can be organized by their transformation properties under the custodial SU(2) symmetry into a fiveplet, a triplet, and two singlets.  The fiveplet and triplet states are given by
\begin{eqnarray}
	&&H_5^{++} = \chi^{++}, \qquad
	H_5^+ = \frac{\left(\chi^+ - \xi^+\right)}{\sqrt{2}}, \qquad
	H_5^0 = \sqrt{\frac{2}{3}} \xi^{0,r} - \sqrt{\frac{1}{3}} \chi^{0,r}, \nonumber \\
	&&H_3^+ = - s_H \phi^+ + c_H \frac{\left(\chi^++\xi^+\right)}{\sqrt{2}}, \qquad
	H_3^0 = - s_H \phi^{0,i} + c_H \chi^{0,i},
\end{eqnarray}
where the vevs are parameterized by
\begin{equation}
	c_H \equiv \cos\theta_H = \frac{v_{\phi}}{v}, \qquad
	s_H \equiv \sin\theta_H = \frac{2\sqrt{2}\,v_\chi}{v},
\end{equation}
and we have decomposed the neutral fields into real and imaginary parts according to
\begin{equation}
	\phi^0 \to \frac{v_{\phi}}{\sqrt{2}} + \frac{\phi^{0,r} + i \phi^{0,i}}{\sqrt{2}},
	\qquad
	\chi^0 \to v_{\chi} + \frac{\chi^{0,r} + i \chi^{0,i}}{\sqrt{2}}, 
	\qquad
	\xi^0 \to v_{\chi} + \xi^{0,r}.
\end{equation}
The masses within each custodial multiplet are degenerate at tree level and can be written (after eliminating $\mu_2^2$ and $\mu_3^2$ in favor of the vevs) as\footnote{Note that the ratio $M_1/v_{\chi}$ can be written using the minimization condition $\partial V/ \partial v_{\chi} = 0$ as
\begin{equation}
	\frac{M_1}{v_{\chi}} = \frac{4}{v_{\phi}^2} 
	\left[ \mu_3^2 + (2 \lambda_2 - \lambda_5) v_{\phi}^2 
	+ 4(\lambda_3 + 3 \lambda_4) v_{\chi}^2 - 6 M_2 v_{\chi} \right],
\end{equation}
which is finite in the limit $v_{\chi} \to 0$.}
\begin{eqnarray}
	m_5^2 &=& \frac{M_1}{4 v_{\chi}} v_\phi^2 + 12 M_2 v_{\chi} 
	+ \frac{3}{2} \lambda_5 v_{\phi}^2 + 8 \lambda_3 v_{\chi}^2, \nonumber \\
	m_3^2 &=&  \frac{M_1}{4 v_{\chi}} (v_\phi^2 + 8 v_{\chi}^2) 
	+ \frac{\lambda_5}{2} (v_{\phi}^2 + 8 v_{\chi}^2) 
	= \left(  \frac{M_1}{4 v_{\chi}} + \frac{\lambda_5}{2} \right) v^2.
\end{eqnarray}

The two custodial-singlet mass eigenstates are given by
\begin{equation}
	h = \cos \alpha \, \phi^{0,r} - \sin \alpha \, H_1^{0\prime},  \qquad
	H = \sin \alpha \, \phi^{0,r} + \cos \alpha \, H_1^{0\prime},
	\label{mh-mH}
\end{equation}
where 
\begin{equation}
	H_1^{0 \prime} = \sqrt{\frac{1}{3}} \xi^{0,r} + \sqrt{\frac{2}{3}} \chi^{0,r},
\end{equation}
and we will use the shorthand $c_{\alpha} \equiv \cos\alpha$, $s_{\alpha} \equiv \sin\alpha$.
The mixing angle $\alpha$ and masses are given by
\begin{eqnarray}
	&&\sin 2 \alpha =  \frac{2 \mathcal{M}^2_{12}}{m_H^2 - m_h^2},    \qquad
	\cos 2 \alpha =  \frac{ \mathcal{M}^2_{22} - \mathcal{M}^2_{11}  }{m_H^2 - m_h^2}, 
	\nonumber \\
	&&m^2_{h,H} = \frac{1}{2} \left[ \mathcal{M}_{11}^2 + \mathcal{M}_{22}^2
	\mp \sqrt{\left( \mathcal{M}_{11}^2 - \mathcal{M}_{22}^2 \right)^2 
	+ 4 \left( \mathcal{M}_{12}^2 \right)^2} \right],
	\label{eq:hmass}
\end{eqnarray}
where we choose $m_h < m_H$, and 
\begin{eqnarray}
	\mathcal{M}_{11}^2 &=& 8 \lambda_1 v_{\phi}^2, \nonumber \\
	\mathcal{M}_{12}^2 &=& \frac{\sqrt{3}}{2} v_{\phi} 
	\left[ - M_1 + 4 \left(2 \lambda_2 - \lambda_5 \right) v_{\chi} \right], \nonumber \\
	\mathcal{M}_{22}^2 &=& \frac{M_1 v_{\phi}^2}{4 v_{\chi}} - 6 M_2 v_{\chi} 
	+ 8 \left( \lambda_3 + 3 \lambda_4 \right) v_{\chi}^2.
\end{eqnarray}

\subsection{H5plane benchmark}

The H5plane benchmark scenario for the GM model was introduced in Ref.~\cite{deFlorian:2016spz}.   It is designed to facilitate LHC searches for $H_5^{\pm\pm}$ and $H_5^{\pm}$ in vector boson fusion with decays to $W^{\pm}W^{\pm}$ and $W^{\pm}Z$, respectively. 
It is specified as in Table~\ref{tab:H5plane}, in a form that is easily implemented in the model calculator {\tt GMCALC}~\cite{Hartling:2014xma}.
After imposing the existing direct search constraints on $H_5^{\pm\pm}$, the benchmark has the following features:
\begin{itemize}
\item It comes close to fully populating the theoretically-allowed region of the $m_5$--$s_H$ plane for $m_5 \in [200,3000]~{\rm GeV}$, as shown in Fig.~\ref{fig:fullscan} (see below).
\item It has $m_3 > m_5$ over the whole benchmark plane, so that the Higgs-to-Higgs decays $H_5 \to H_3 H_3$ and $H_5 \to H_3 V$ are kinematically forbidden, leaving only the decays $H_5 \to VV$ at tree level; i.e., ${\rm BR}(H_5 \to VV) = 1$.
\item The entire benchmark satisfies indirect constraints from $B$ physics, the most stringent of which is $b \to s \gamma$~\cite{Hartling:2014aga}.
\item The region still allowed by direct searches is currently unconstrained by LHC measurements of the couplings of the 125~GeV Higgs boson, as we will show in this paper.
\end{itemize}

\begin{table}
\begin{tabular}{lll}
\hline \hline
Fixed parameters & Variable parameters & Dependent parameters \\
\hline
$G_F = 1.1663787 \times 10^{-5}$~GeV$^{-2}$ & $m_5 \in [200, 3000]$~GeV & $\lambda_2 = 0.4(m_5/1000~{\rm GeV})$ \\
$m_h = 125$~GeV & $s_H \in (0,1)$ & $M_1 = \sqrt{2} s_H (m_5^2 + v^2)/v$ \\
$\lambda_3 = -0.1$ & & $M_2 = M_1/6$ \\
$\lambda_4 = 0.2$ & & \\
\hline\hline
\end{tabular}
\caption{Specification of the H5plane benchmark scenario for the Georgi-Machacek model.  These input parameters correspond to {\tt INPUTSET = 4} in {\tt GMCALC}~\cite{Hartling:2014xma}.}
\label{tab:H5plane}
\end{table}

In {\tt INPUTSET = 4} of {\tt GMCALC}, the nine parameters of the scalar potential in Eq.~(\ref{eq:potential}) are fixed in terms of the nine input parameters $m_h$, $m_5$, $s_H$,
$\lambda_2$, $\lambda_3$, $\lambda_4$, $M_1$, $M_2$, and $v = (\sqrt{2} G_F)^{-1/2}$.  The quartic coupling $\lambda_5$ is computed from these using
\begin{equation}
	\lambda_5 = \frac{2 m_5^2}{3 c_H^2 v^2} - \frac{\sqrt{2} M_1}{3 s_H v}
                - \frac{2\sqrt{2} M_2 \, s_H}{c_H^2 v} - \frac{2 \lambda_3 \, s_H^2}{3 c_H^2}.
	\label{eqn:lambda5}
\end{equation}
The quartic coupling $\lambda_1$ (which depends on $\lambda_5$) is computed using
\begin{equation}
	\lambda_1 = \frac{1}{8 c_H^2 v^2} \left\{ m_h^2 + \frac{3 c_H^2 v^2 \left[ -M_1 + \sqrt{2}(2\lambda_2-\lambda_5)s_H v \right]^2}
                 {2\sqrt{2}M_1\frac{c_H^2}{s_H}v - 6\sqrt{2} M_2 s_H v + 4(\lambda_3+3\lambda_4)s_H^2 v^2 - 4m_h^2} \right\}.
	\label{eqn:lambda1}
\end{equation}
The mass-squared parameter $\mu_2^2$ (which depends on $\lambda_1$ and $\lambda_5$) is computed using
\begin{equation}
	\mu_2^2 = -4\lambda_1 c_H^2 v^2 - \frac{3}{8}(2\lambda_2-\lambda_5)s_H^2 v^2 
		+ \frac{3\sqrt{2}}{8}M_1 s_H v,
	\label{eqn:mu2sq}
\end{equation}
and $\mu_3^2$ is computed using
\begin{equation}\label{eqn:mu3sq}
 \mu_3^2 = \frac{2}{3} m_5^2 + \frac{\sqrt{2}M_1 c_H^2 v}{6s_H} - 2\lambda_2 c_H^2 v^2
             - \frac{1}{6}(7\lambda_3 + 9\lambda_4)s_H^2 v^2 - \frac{\sqrt{2}}{2}M_2 s_H v.
\end{equation}

In Fig.~\ref{fig:fullscan} we show the allowed region in the $m_5$--$s_H$ plane for the full GM model (red points) and the allowed region for the H5plane benchmark scenario (entire region below both the black and blue curves), as generated using {\tt GMCALC~1.2.1} with $m_h = 125$~GeV.  In both cases we impose the theoretical constraints from perturbative unitarity of the scalar quartic couplings, bounded-from-belowness of the scalar potential, and the absence of deeper alternative minima, as described in Ref.~\cite{Hartling:2014zca}, as well as the indirect constraints from $b \to s \gamma$ and the $S$ parameter following Ref.~\cite{Hartling:2014aga} (we use the ``loose'' constraint on $b \to s \gamma$ as described in Ref.~\cite{Hartling:2014aga}); all of these constraints are implemented in {\tt GMCALC}.  We also impose the direct experimental constraint from a CMS search for $H_5^{\pm\pm}$~\cite{Khachatryan:2014sta} (described in more detail below), which excludes the area above the blue curve in the context of the H5plane benchmark.  The red points represent a scan over the full GM model parameter space.  The entire area below the black curve (obtained by scanning $m_5$ and $s_H$ in the H5plane benchmark) represents the theoretically-allowed region in the H5plane benchmark: as advertised, it nearly, but not quite entirely, populates the entire range of $s_H$ that is accessible in the full GM model for any given value of $m_5$ between 200 and 3000~GeV.  This makes the H5plane scenario a good benchmark for the interpretation of searches for $H_5^{\pm}$ and $H_5^{\pm\pm}$ in vector boson fusion, for which the signal rate and kinematics depend only on $m_5$, $s_H$, and the $H_5$ branching ratios into vector boson pairs.  We note however that the accessible ranges of other observables are not necessarily fully populated by the H5plane benchmark; this will be particularly dramatic for the mass splittings among the heavy Higgs bosons.

\begin{figure}
\resizebox{\textwidth}{!}{\includegraphics{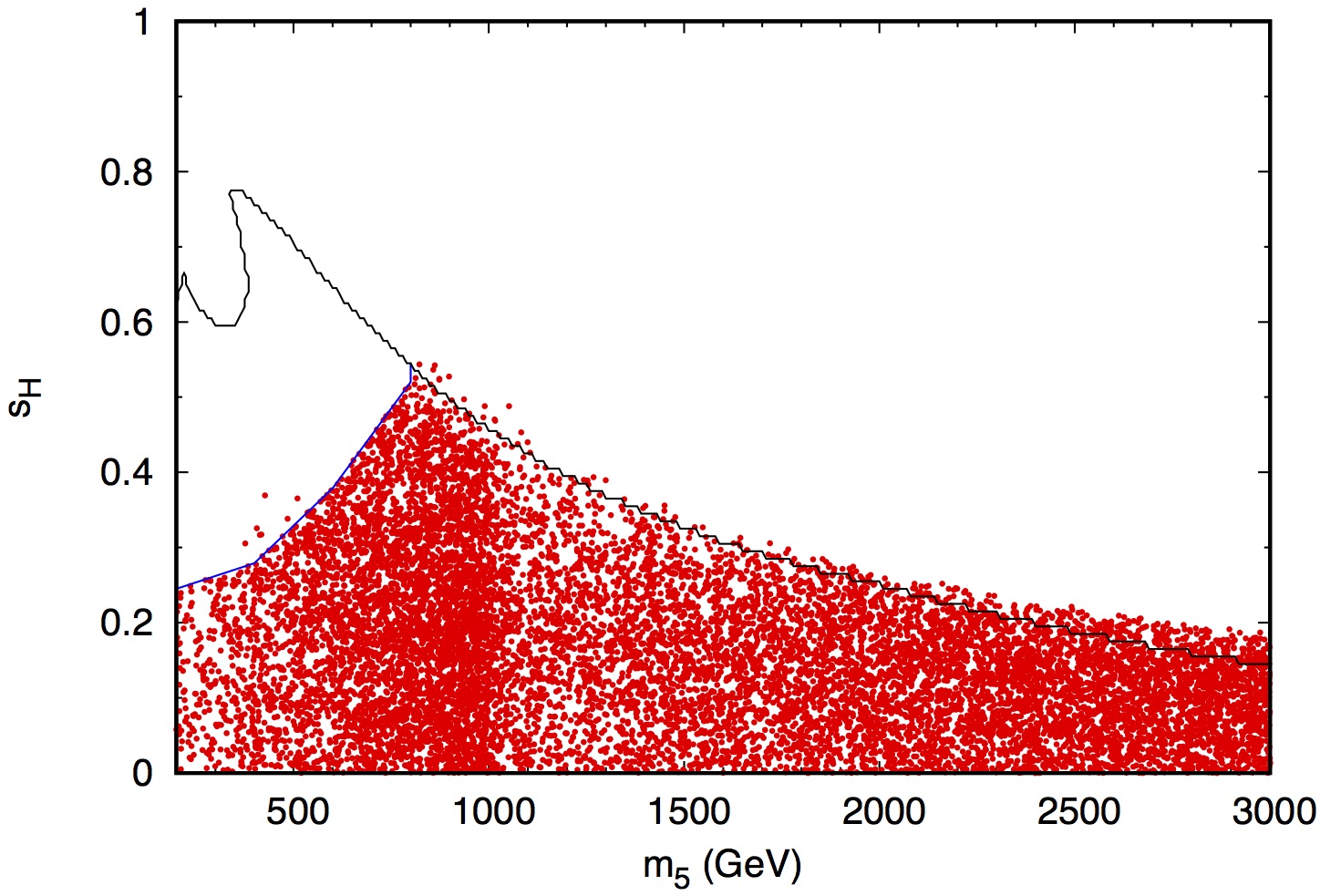}}
\caption{Theoretically and experimentally allowed parameter region in the $m_5$--$s_H$ plane in the H5plane benchmark (entire region below both the black and blue curves) and the full GM model (red points).  The black curve delimits the region allowed by theoretical constraints in the H5plane benchmark and the blue curve represents the upper bound on $s_H$ from a direct search for $H_5^{\pm\pm}$ from Ref.~\cite{Khachatryan:2014sta}. See text for details.}
\label{fig:fullscan}
\end{figure}

The CMS search in Ref.~\cite{Khachatryan:2014sta} currently provides the most stringent direct experimental constraint on the GM model for $m_5$ above 200~GeV.\footnote{For comparison, the 95\% confidence level constraint obtained in Ref.~\cite{Chiang:2014bia} from an ATLAS measurement of the cross section for like-sign $W$ boson pairs in vector boson fusion~\cite{Aad:2014zda} excludes $s_H$ values above 0.39 for $m_5 = 200$~GeV, rising to 0.74 for $m_5 = 600$~GeV.  LHC searches for $H_5^{\pm}$ in the $WZ$ final state~\cite{Aad:2015nfa,Sirunyan:2017sbn} are currently slightly less constraining than the search for $H_5^{\pm\pm}$.}  
This search looked for a doubly-charged scalar produced in vector boson fusion (VBF) and decaying to two like-sign $W$ bosons which in turn decay leptonically, using 19.4~fb$^{-1}$ of proton-proton collision data at a centre-of-mass energy of 8~TeV.  This search set a 95\% confidence level upper bound on the cross section times branching ratio, $\sigma({\rm VBF} \to H^{\pm\pm}) \times {\rm BR}(H^{\pm\pm} \to W^{\pm}W^{\pm})$, as a function of the doubly-charged Higgs boson mass.  The H5plane benchmark is designed so that ${\rm BR}(H_5^{\pm\pm} \to W^{\pm} W^{\pm}) = 1$, so that the CMS constraint becomes an upper bound on the cross section $\sigma({\rm VBF} \to H_5^{\pm\pm})$, which is proportional to $s_H^2$.  We translated this into an upper bound on $s_H$ in the H5plane benchmark using the ${\rm VBF} \to H_5^{\pm\pm}$ cross sections calculated for the 8~TeV LHC at next-to-next-to-leading order in QCD in Ref.~\cite{Zaro:2015ika} (we did not take into account the theoretical uncertainties in these predictions in computing the limit).  This constraint in the H5plane benchmark is shown as the blue curve in Fig.~\ref{fig:fullscan}; when combined with the theoretical constraints, it limits $s_H < 0.55$ in the H5plane benchmark.  In a full scan of the GM model, some allowed points appear that have ${\rm BR}(H_5^{\pm\pm} \to W^{\pm} W^{\pm}) < 1$, because decays into $H_3^{\pm} W^{\pm}$ are kinematically allowed.  Since the CMS constraint applies to the product $\sigma({\rm VBF} \to H_5^{\pm\pm}) \times {\rm BR}(H_5^{\pm\pm} \to W^{\pm}W^{\pm})$, this results in a few of the allowed red points in Fig.~\ref{fig:fullscan} falling above the blue curve.  The number of such points is quite small, though, because most points in the full GM model scan that have ${\rm BR}(H_5^{\pm\pm} \to W^{\pm} W^{\pm}) < 1$ also have small $s_H$, putting them below the blue curve anyway.

\section{Properties of the H5plane benchmark}
\label{sec:properties}

\subsection{Decays of $H_5$}

The H5plane benchmark was designed so that $m_3 > m_5$ over the entire benchmark plane, so that the decay $H_5^{\pm\pm} \to W^{\pm} W^{\pm}$ is the only kinematically-allowed decay for $H_5^{\pm\pm}$.  This makes direct searches for $H_5^{\pm\pm}$ in the $WW$ final state particularly easy to interpret.  Decays of $H_5^{\pm}$ to $W^{\pm}Z$ are then also the only kinematically-accessible tree-level decay of $H_5^{\pm}$ (the loop-induced decay $H_5^{\pm} \to W^{\pm}\gamma$ is allowed, but has a very small branching ratio for $m_5 \geq 200$~GeV), so that direct searches for the singly-charged state in this final state are also easy to interpret.  This was used in the GM model interpretation of the ATLAS and CMS searches for $H_5^{\pm}$ in Refs.~\cite{Aad:2015nfa,Sirunyan:2017sbn} (these searches are less constraining on the GM model parameter space than that of Ref.~\cite{Khachatryan:2014sta}).

\begin{figure}
\resizebox{0.5\textwidth}{!}{\includegraphics{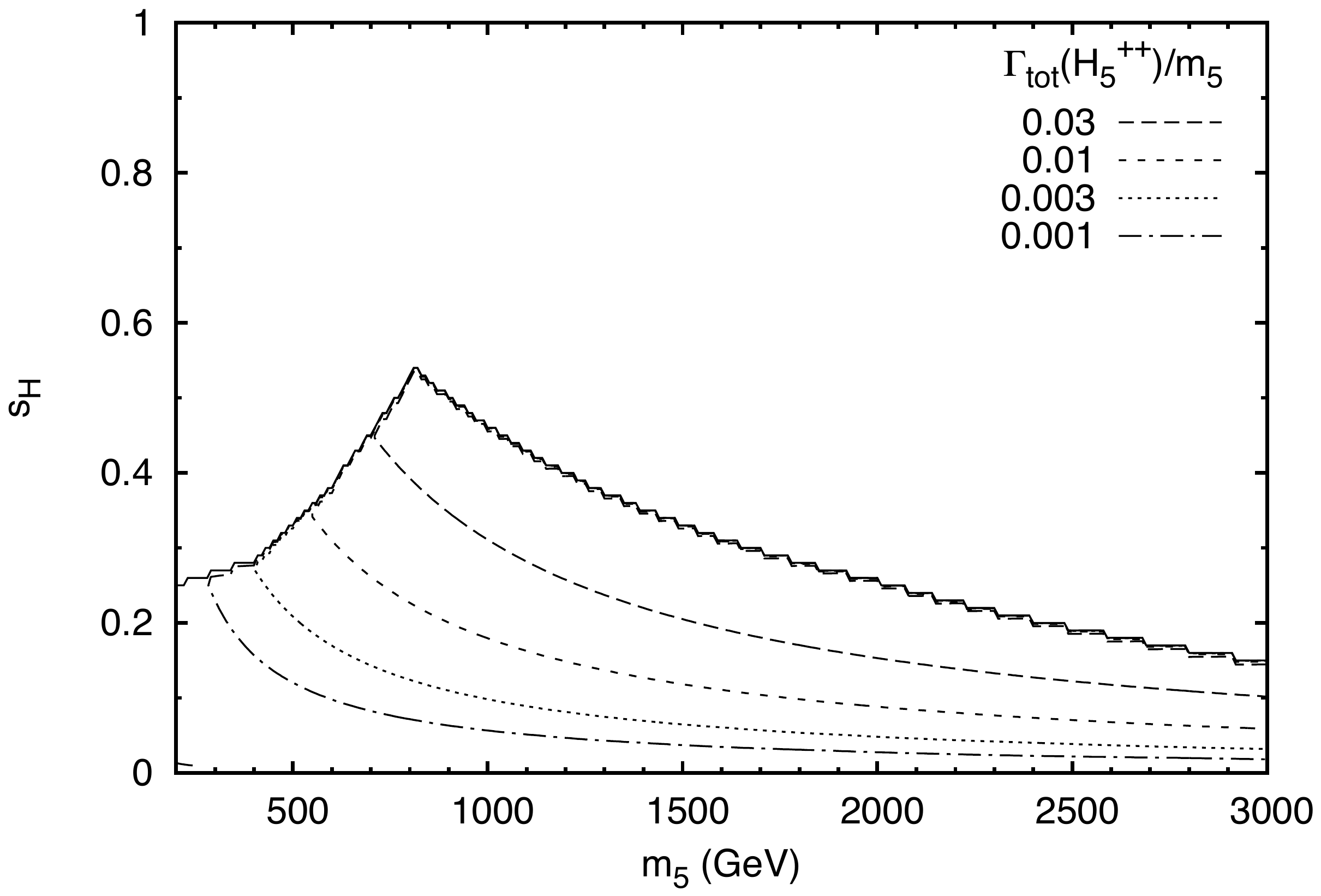}}%
\resizebox{0.5\textwidth}{!}{\includegraphics{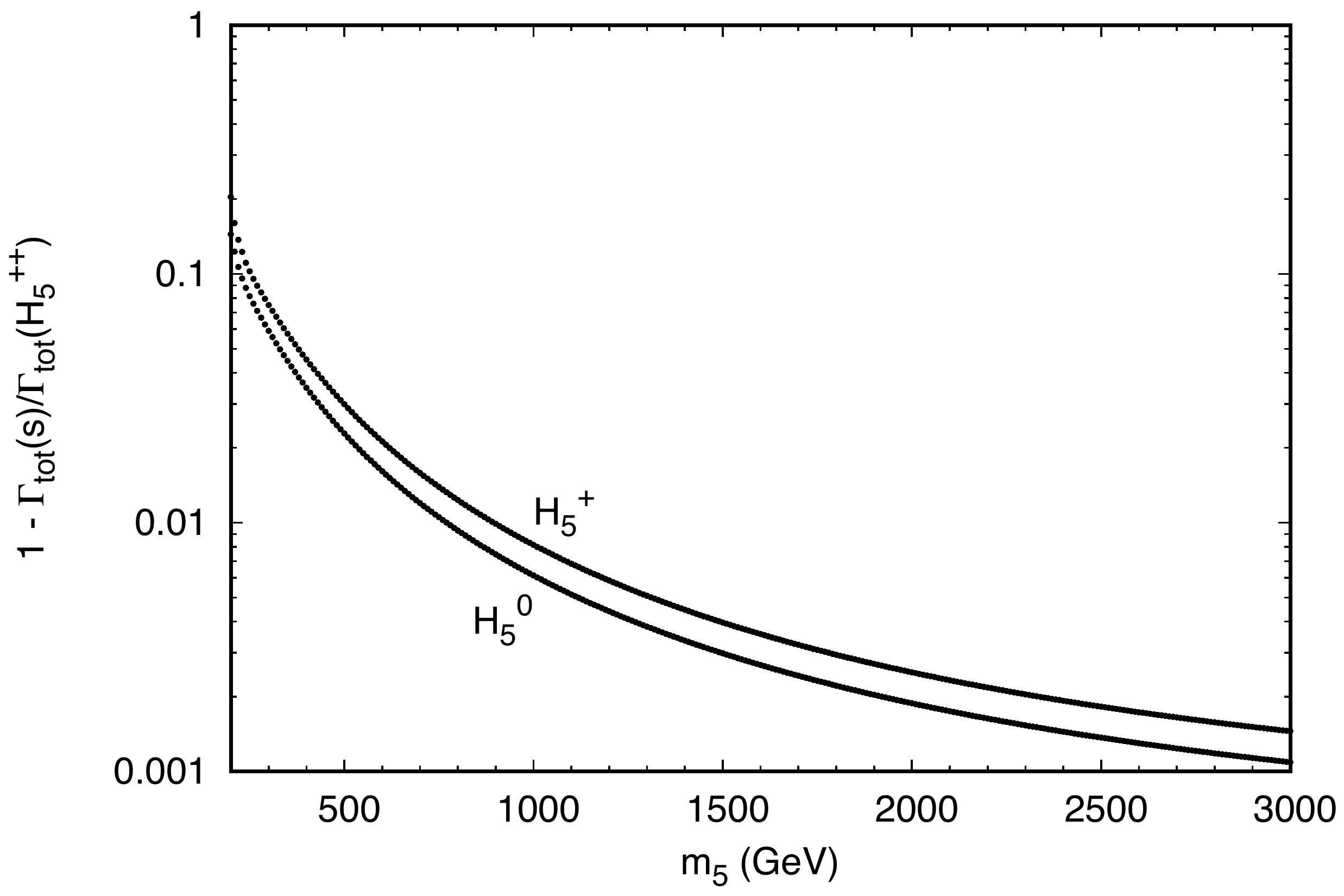}}
\caption{Left: Contours of $\Gamma_\text{tot} / m_5$ for $H_5^{++}$ in the GM model H5plane benchmark.  The value of $\Gamma_\text{tot} / m_5$ reaches a maximum of 0.08 along the upper boundary of the allowed region for $m_5 \gtrsim 800$~GeV, and goes to zero at $s_H = 0$.
Right: Deviation from unity of the ratio of total widths of scalar $s = H_5^+$ and $H_5^0$ to that of $H_5^{++}$ as a function of $m_5$ in the H5plane benchmark.  Direct constraints from a CMS search for $H_5^{\pm\pm} \to W^{\pm}W^{\pm}$ in vector boson fusion~\cite{Khachatryan:2014sta} have been applied.}
\label{fig:h5pp_dir_data}
\end{figure}

In the left panel of Fig.~\ref{fig:h5pp_dir_data} we show the total width of $H_5^{\pm\pm}$ normalized to its mass.  This width-to-mass ratio reaches a maximum of 8\% for the largest theoretically-allowed values of $s_H$ when $m_5 > 800$~GeV.  The right panel of Fig.~\ref{fig:h5pp_dir_data} shows the deviation from unity of the ratio of partial widths of $H_5^{\pm}$ and $H_5^0$ divided by that of $H_5^{\pm\pm}$ as a function of $m_5$.  These ratios are independent of $s_H$.  The widths of $H_5^+$ and $H_5^0$ are about 10\% smaller than that of $H_5^{++}$ for $m_5 \sim 200$~GeV, with the difference decreasing to less than 1\% for $m_5 \gtrsim 1000$~GeV.  In the H5plane benchmark, this width difference is solely due to the kinematic effect of the different masses of the $WW$, $WZ$, and $ZZ$ final states.

\subsection{$H_3$--$H_5$ mass splitting}

\begin{figure}
\resizebox{0.5\textwidth}{!}{\includegraphics{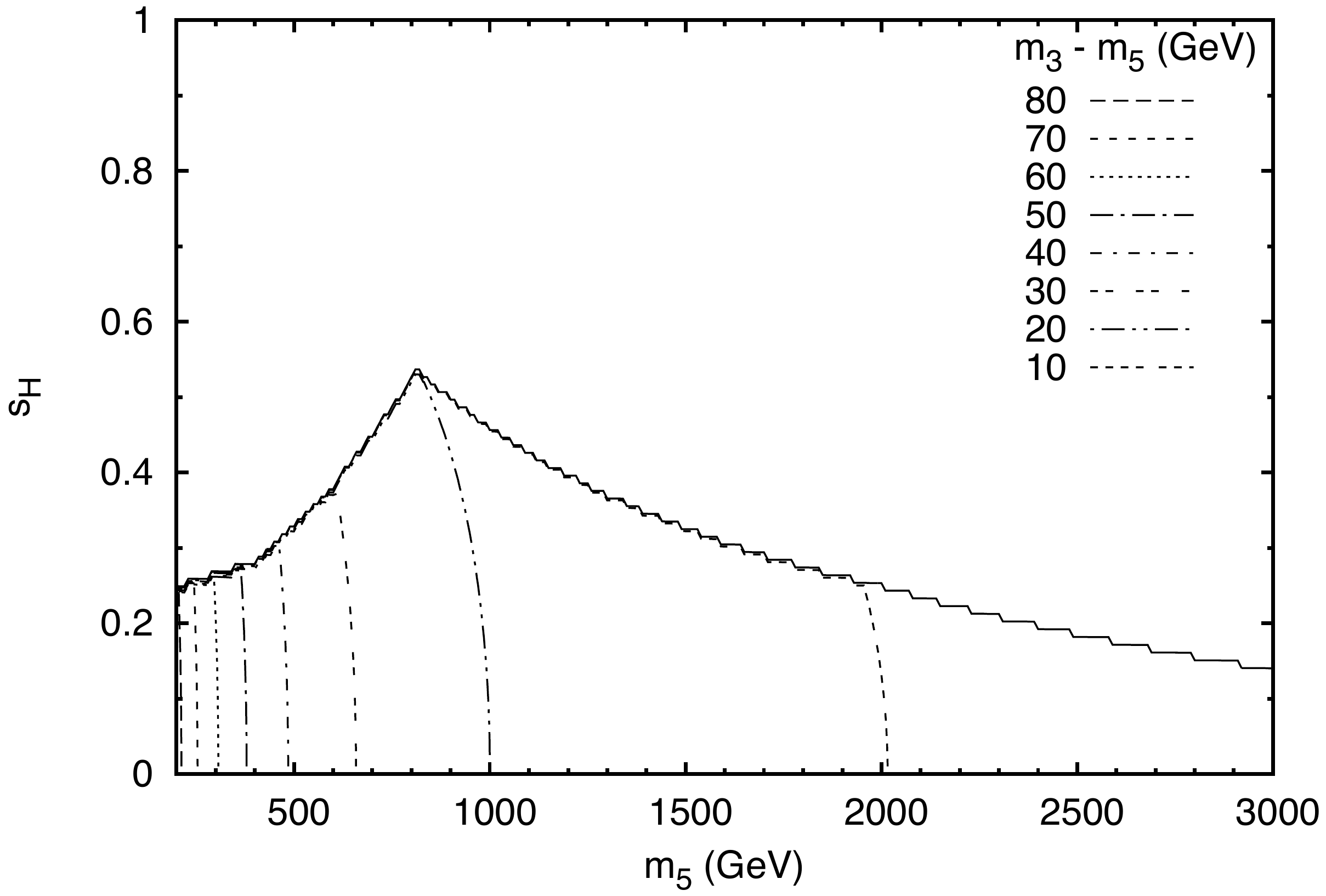}}%
\resizebox{0.5\textwidth}{!}{\includegraphics{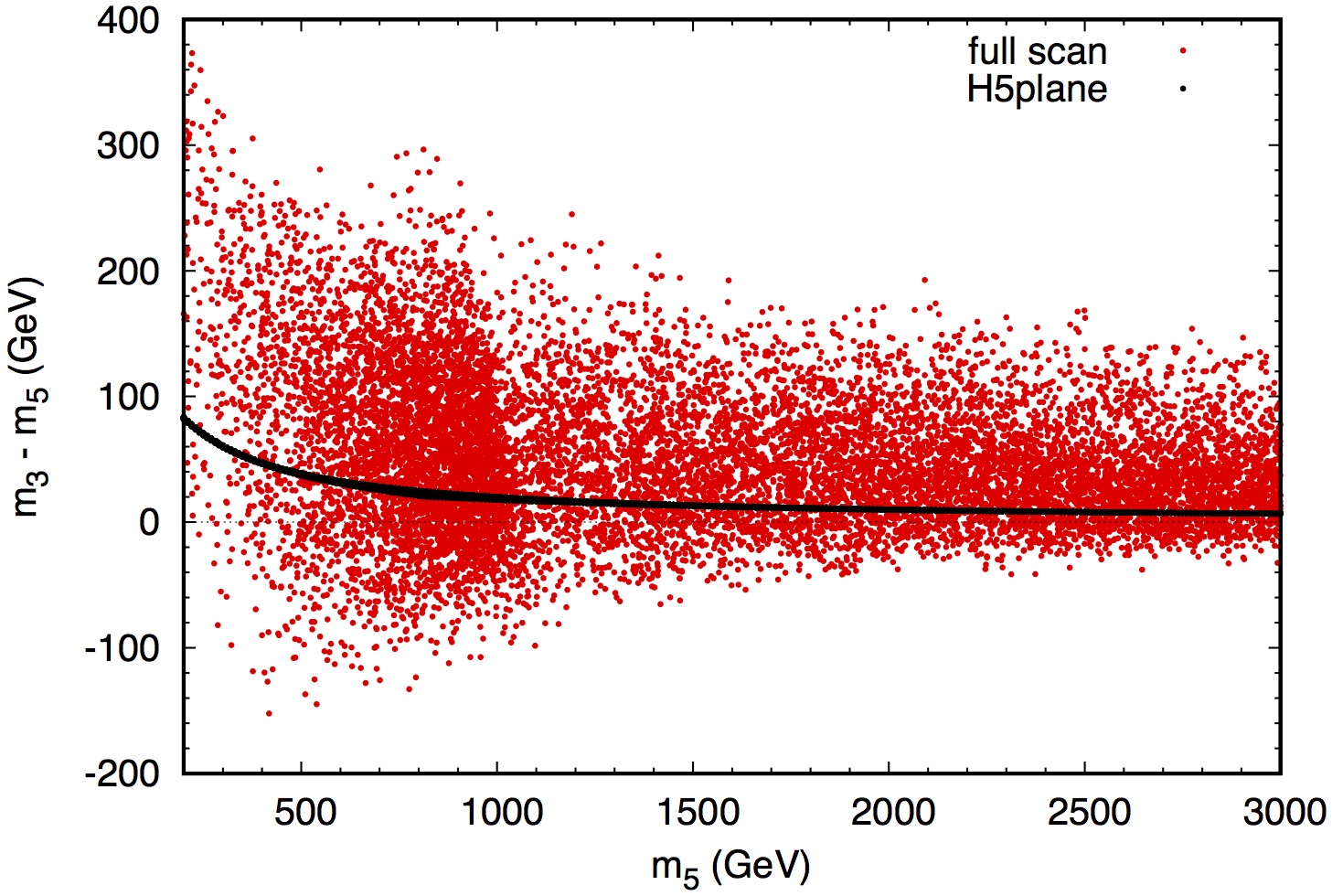}}
\caption{Left: Contours of $m_3-m_5$ in the H5plane benchmark.  The value of $m_3 - m_5$ ranges from 6.7~GeV to 84~GeV.
Right: Mass difference $m_3 - m_5$ as a function of $m_5$ in the H5plane benchmark (black points) and in a full scan of the GM model parameter space (red points).  Indirect constraints from $b \to s \gamma$ and the $S$ parameter~\cite{Hartling:2014aga} and direct constraints from a CMS search for $H_5^{\pm\pm} \to W^{\pm}W^{\pm}$ in vector boson fusion~\cite{Khachatryan:2014sta} have been applied.}
\label{fig:mh3minusmh5_data}
\end{figure}

In the left panel of Fig.~\ref{fig:mh3minusmh5_data} we show the mass splitting $m_3 - m_5$ in the H5plane benchmark.  This splitting depends mainly on $m_5$, and varies from 84~GeV at $m_5 = 200$~GeV to about 7~GeV at $m_5 = 3000$~GeV.  In the right panel of Fig.~\ref{fig:mh3minusmh5_data} we plot $m_3-m_5$ as a function of $m_5$ scanning over all the other free parameters in the H5plane benchmark (black points) and the full GM model (red points), where we have imposed the indirect constraints from $b \to s \gamma$ and the $S$ parameter~\cite{Hartling:2014aga} and direct constraints from the CMS search for $H_5^{\pm\pm} \to W^{\pm}W^{\pm}$ in vector boson fusion~\cite{Khachatryan:2014sta}.  It is clear that the variation in the mass difference $m_3 - m_5$ is much greater in the full model scan than it is in the H5plane benchmark.  We can understand this as follows.

The difference between $m_3^2$ and $m_5^2$ can be written in the full GM model as
\begin{equation}
	m_3^2 - m_5^2 = (M_1 - 6M_2) \frac{s_H v}{\sqrt{2}}
                   + \left[ \lambda_5\left(\frac{1}{2}s_H^2-c_H^2\right) - \lambda_3 s_H^2\right] v^2.
	\label{eqn:m32-m52}
\end{equation}
In the H5plane benchmark, the parameter relations simplify this down to
\begin{equation}
	m_3^2-m_5^2 = (m_3 - m_5)(m_3 + m_5) = \left( \frac{2}{3} - \frac{0.3 s_H^2}{c_H^2} \right) v^2.
	\label{eqn:m32-m52_bench}
\end{equation}
The variation of this expression with $s_H$ is fairly minimal: $m_3^2-m_5^2$ changes by less than 10\% between $s_H=0$ and $s_H=0.4$.  This leads to the very narrow range of $m_3-m_5$ covered by the H5plane benchmark scan (black points) in the right panel of Fig.~\ref{fig:mh3minusmh5_data}.  Solving Eq.~(\ref{eqn:m32-m52_bench}) for $m_3-m_5$, the dependence on $m_5$ is due only to a factor of $1/(m_3+m_5) \simeq 1/(2 m_5)$.

In contrast, in the full GM model scan (red points in the right panel of Fig.~\ref{fig:mh3minusmh5_data}), $m_3-m_5$ varies by hundreds of GeV.  This is mostly due to the term proportional to $(M_1 - 6 M_2)$ in Eq.~(\ref{eqn:m32-m52}), which is zero in the H5plane benchmark due to the choice $M_2 = M_1/6$, and the term $-\lambda_5 c_H^2 v^2$, which is not suppressed at small $s_H$.  In the full GM model, $\lambda_5$ can vary between $-8 \pi/3$ and $+8\pi/3$~\cite{Hartling:2014zca}, while in the H5plane benchmark Eq.~(\ref{eqn:lambda5}) reduces to 
\begin{equation}
	\lambda_5 = - \frac{2}{3 c_H^2} (1 - 0.1 s_H^2),
\end{equation}
so that the term $-\lambda_5 c_H^2 v^2$ varies from $2v^2/3$ by less than 2\% for $s_H$ between zero and 0.4 in the H5plane benchmark.  The preference for positive values of $m_3-m_5$ in the full GM model scan is due to the interplay of the theoretical constraints on the model parameters and is apparent already in Fig.~3 of Ref.~\cite{Hartling:2014aga}.  Viable mass spectra in the full GM model, and their implications for cascade decays of the heavier Higgs bosons, have previously been studied in Ref.~\cite{Chiang:2015amq}.

\subsection{Couplings and decays of $h$}

The tree-level couplings of the 125~GeV Higgs boson $h$ in the GM model are given in terms of the underlying parameters by
\begin{equation}
	\kappa^h_f = \frac{c_{\alpha}}{c_H}, \qquad \qquad
	\kappa^h_V = c_{\alpha} c_H - \sqrt{\frac{8}{3}} s_{\alpha} s_H,
\end{equation}
where $\kappa$ is defined in the usual way as the ratio of the coupling in the GM model to the corresponding coupling of the SM Higgs boson~\cite{LHCHiggsCrossSectionWorkingGroup:2012nn}.

\begin{figure}
\resizebox{0.5\textwidth}{!}{\includegraphics{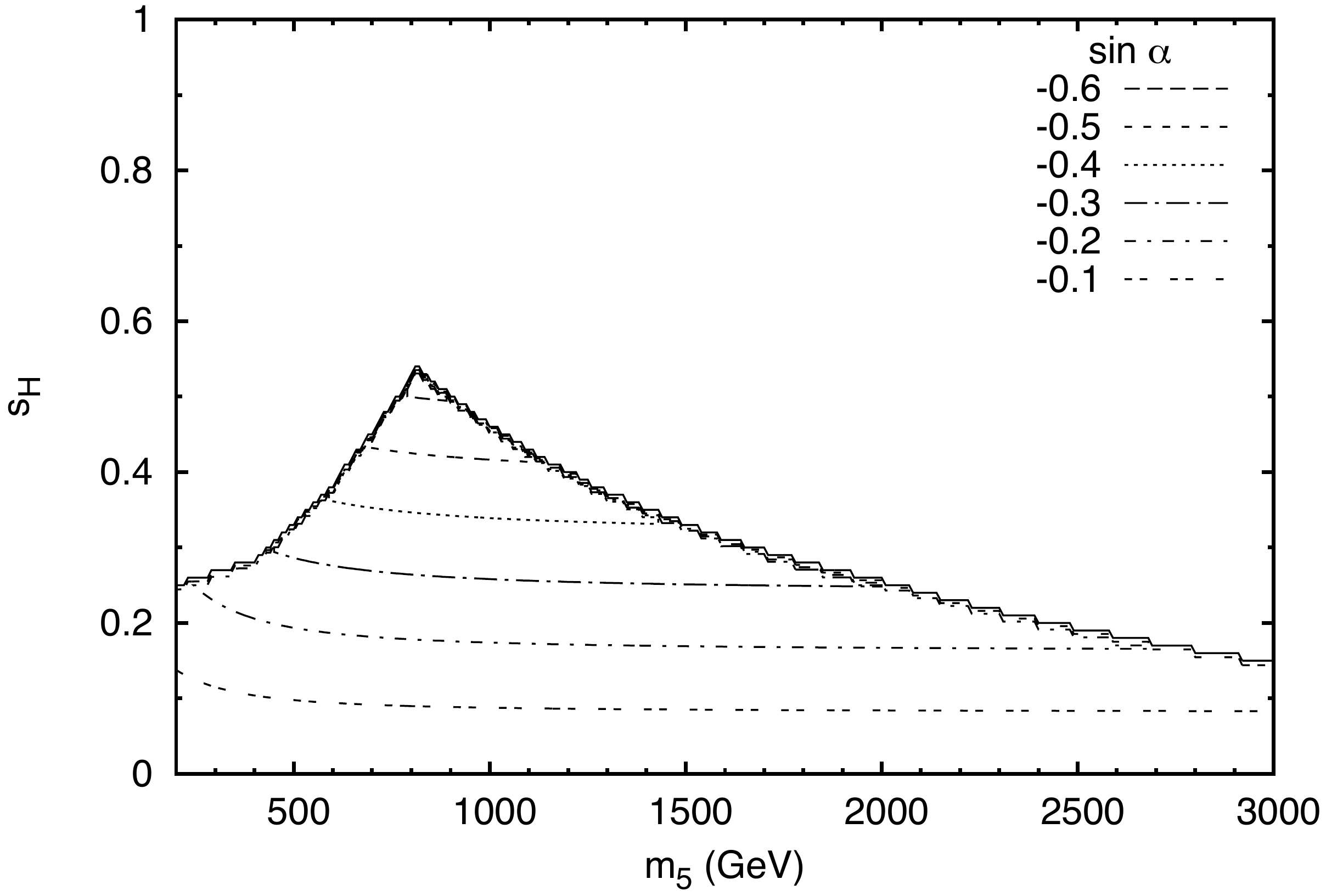}}%
\resizebox{0.5\textwidth}{!}{\includegraphics{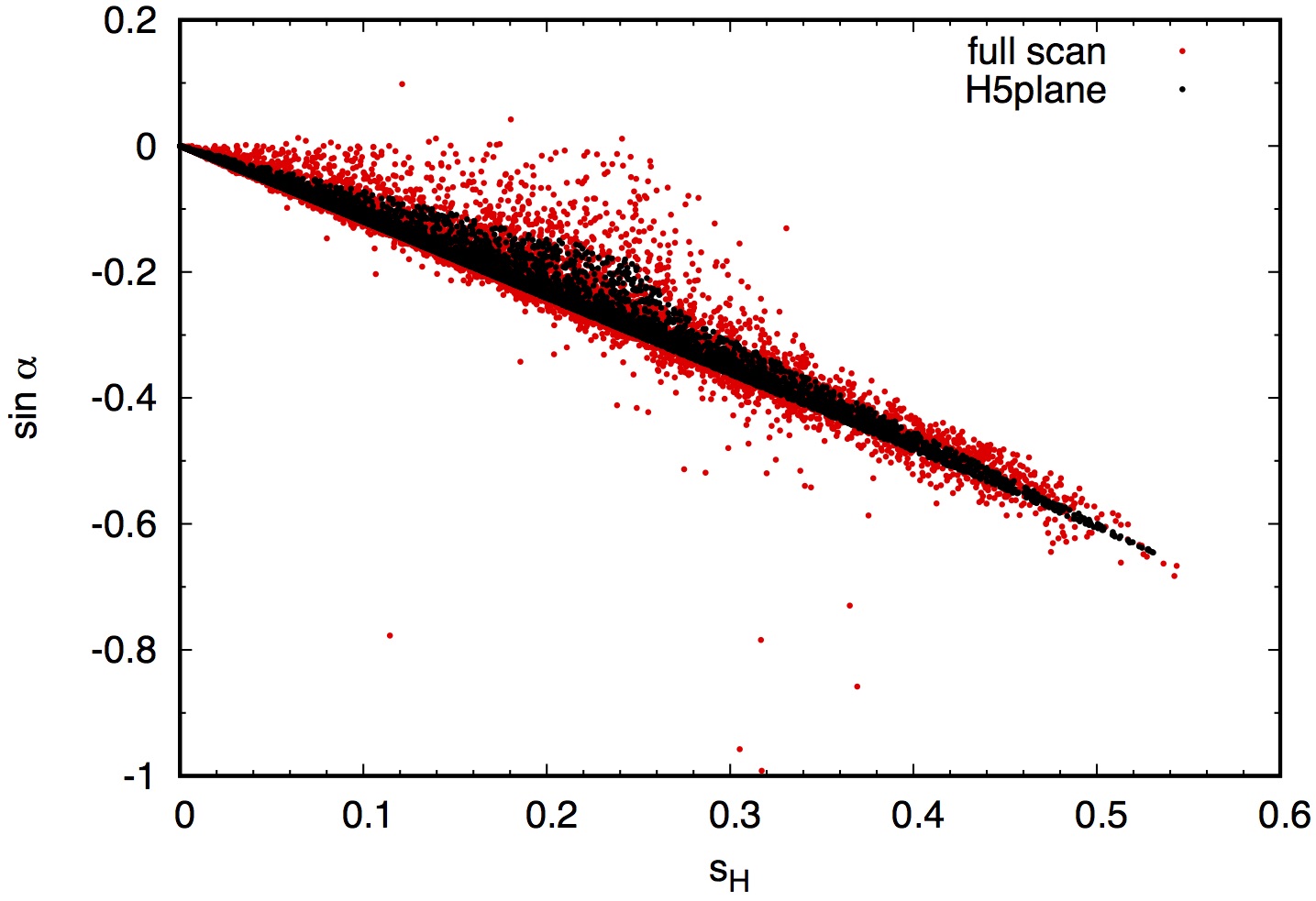}}
\caption{Left: Contours of $\sin \alpha$ in the H5plane benchmark.  The value of $\sin \alpha$ varies between $-0.64$ and $0$.
Right: Correlation between $\sin\alpha$ and $s_H$ in the H5plane benchmark (black points) and in a general GM model scan with $m_5 \geq 200$~GeV (red points).  Indirect constraints from $b \to s \gamma$ and the $S$ parameter~\cite{Hartling:2014aga} and direct constraints from a CMS search for $H_5^{\pm\pm} \to W^{\pm}W^{\pm}$ in vector boson fusion~\cite{Khachatryan:2014sta} have been applied.}
\label{fig:sina_data}
\end{figure}

We first illustrate the variation of the custodial-singlet scalar mixing angle $\sin\alpha$ over the H5plane benchmark in the left panel of Fig.~\ref{fig:sina_data}.  $\sin\alpha$ varies between zero and $-0.64$ in the H5plane benchmark. It is strongly correlated with $s_H$, as shown in the right panel of Fig.~\ref{fig:sina_data}.  This correlation also appears in a full scan of the GM model (red points in the right panel of Fig.~\ref{fig:sina_data}), but is stronger in the H5plane benchmark (black points).

\begin{figure}
\resizebox{0.5\textwidth}{!}{\includegraphics{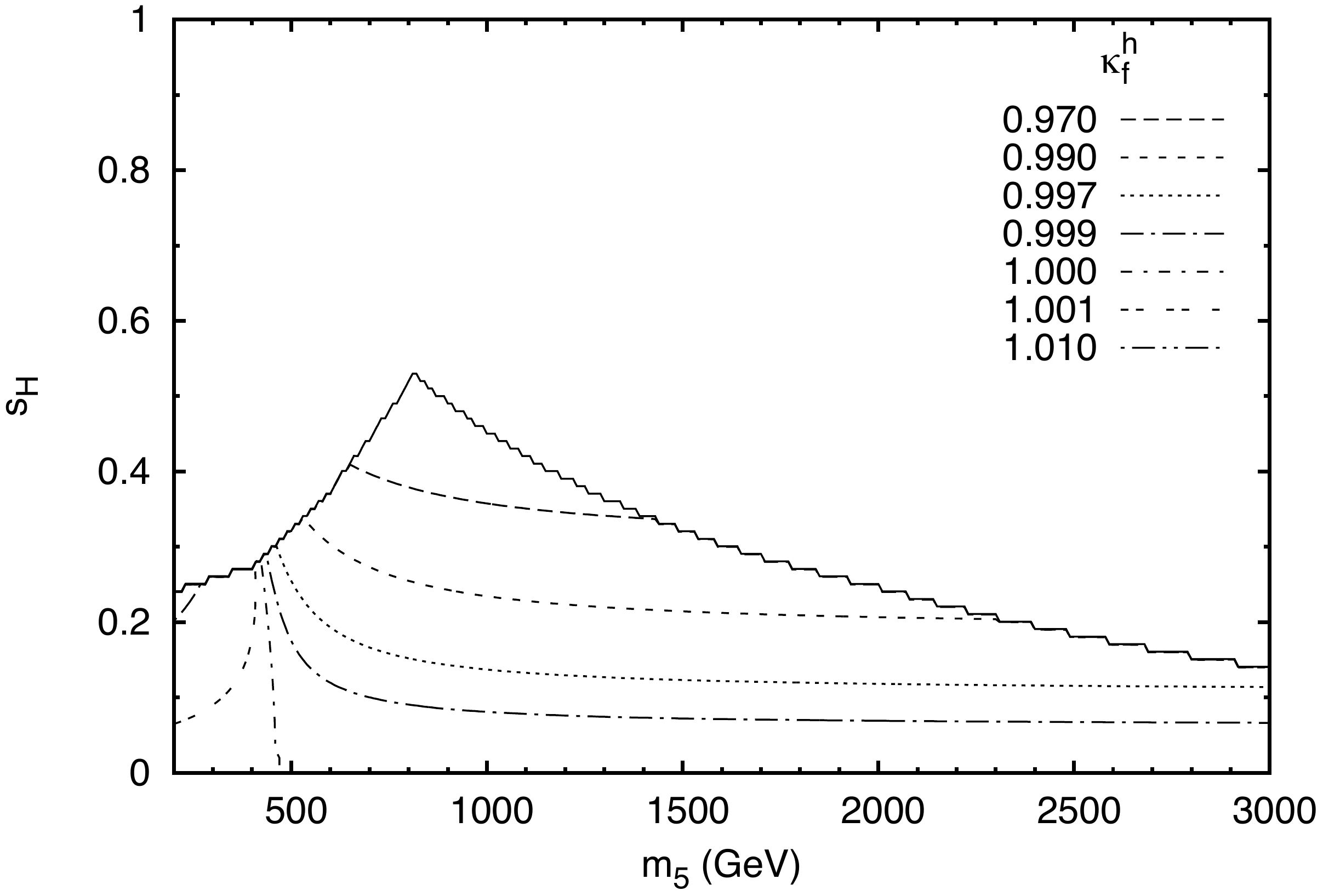}}%
\resizebox{0.5\textwidth}{!}{\includegraphics{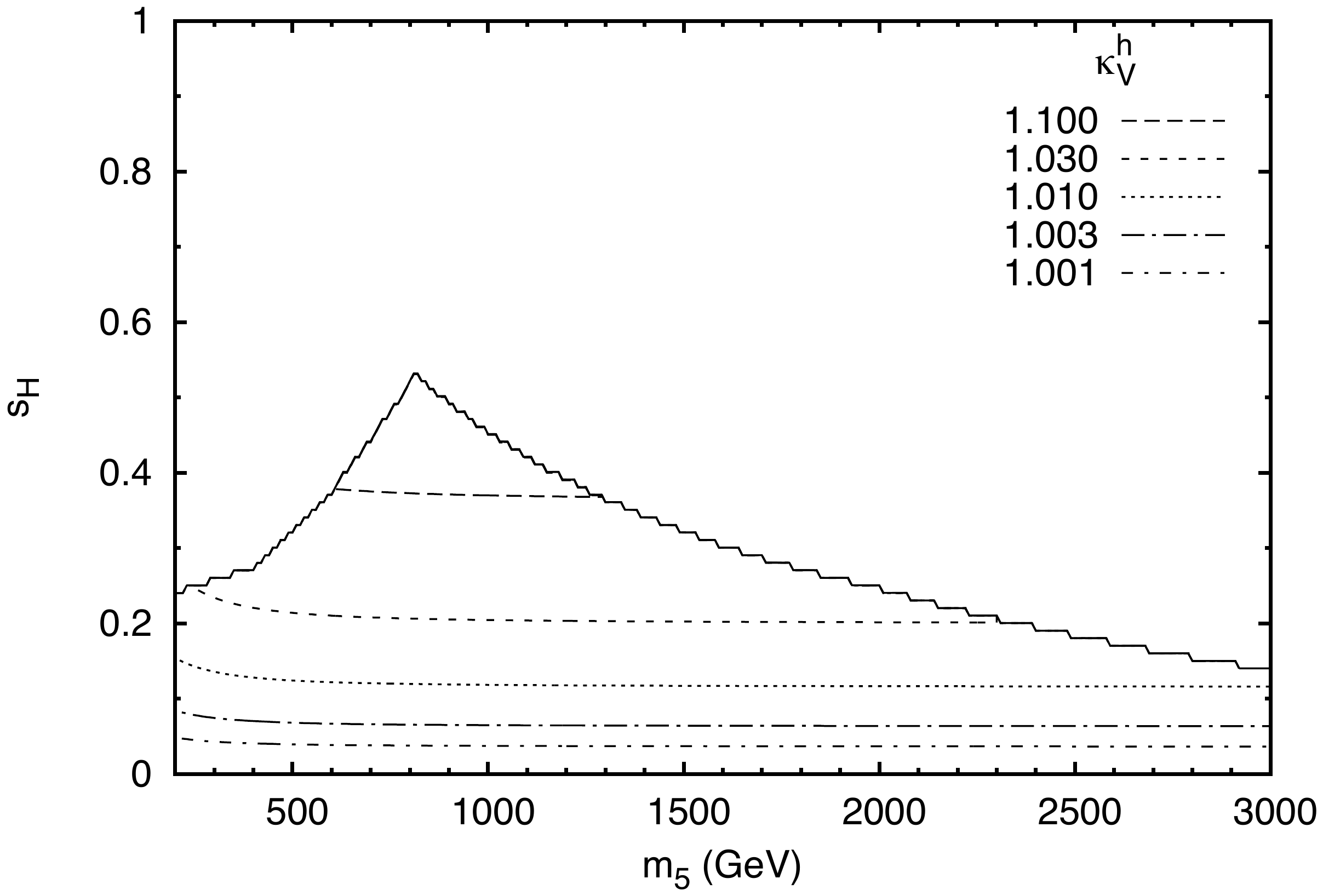}}
\caption{Left: Contours of $\kappa_f^h$ in the H5plane benchmark.
         The value of $\kappa_f^h$ ranges from $0.902$ to $1.014$.
Right: Contours of $\kappa_V^h$ in the H5plane benchmark.
         The value of $\kappa_V^h$ ranges from $1.00$ to $1.21$.}
\label{fig:KVL_data}
\end{figure}

In Fig.~\ref{fig:KVL_data} we plot $\kappa_f^h$ (left panel) and $\kappa_V^h$ (right panel) in the H5plane benchmark.  These couplings remain reasonably close to their SM value of 1 everywhere in the benchmark plane.  The coupling of $h$ to fermions $\kappa_f^h$ varies between 0.902 and 1.014, reaching its smallest values when $s_H$ is large, and the coupling of $h$ to vector bosons $\kappa_V^h$ varies between 1 and 1.21, reaching its largest values when $s_H$ is large.

\begin{figure}
\resizebox{0.5\textwidth}{!}{\includegraphics{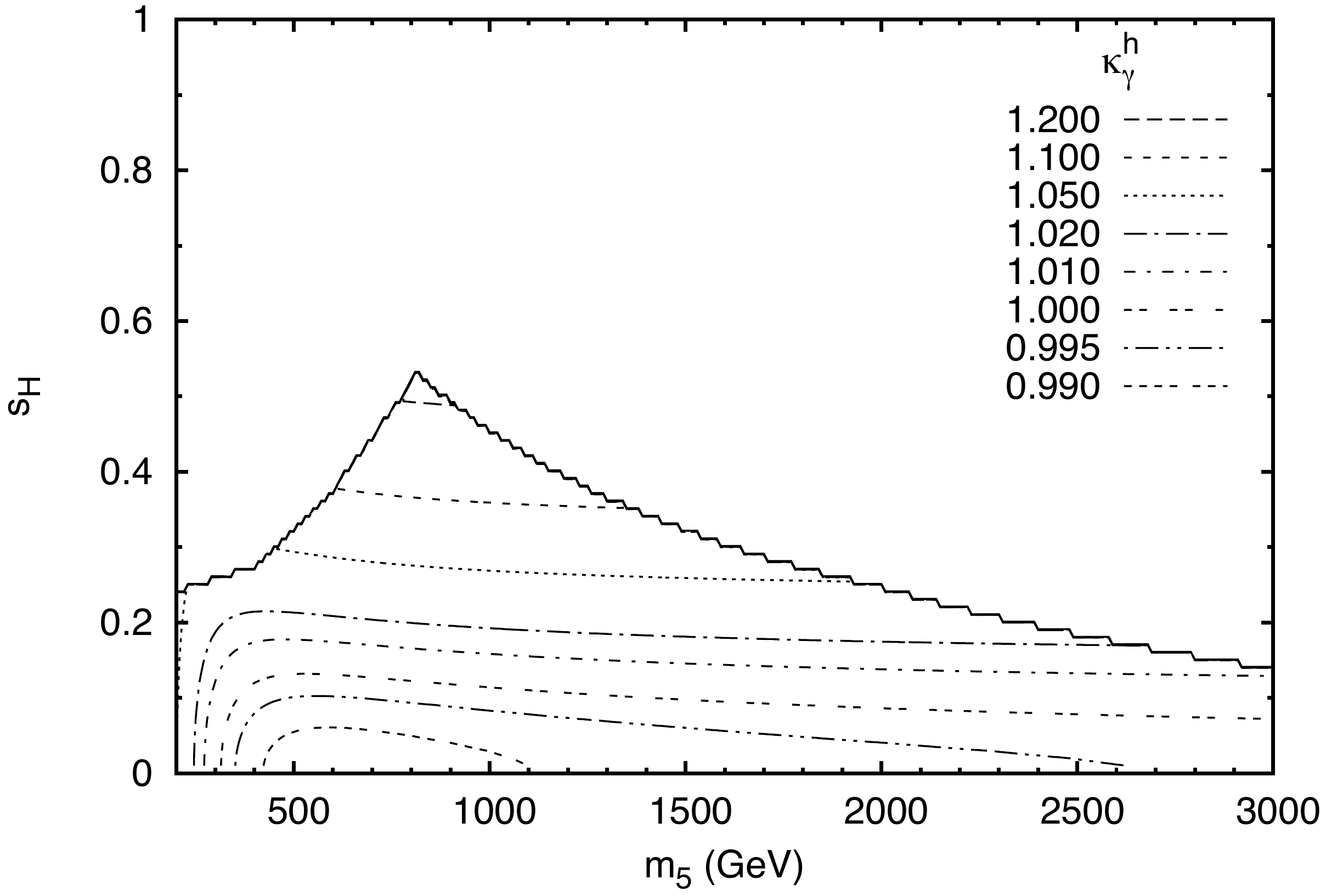}}%
\resizebox{0.5\textwidth}{!}{\includegraphics{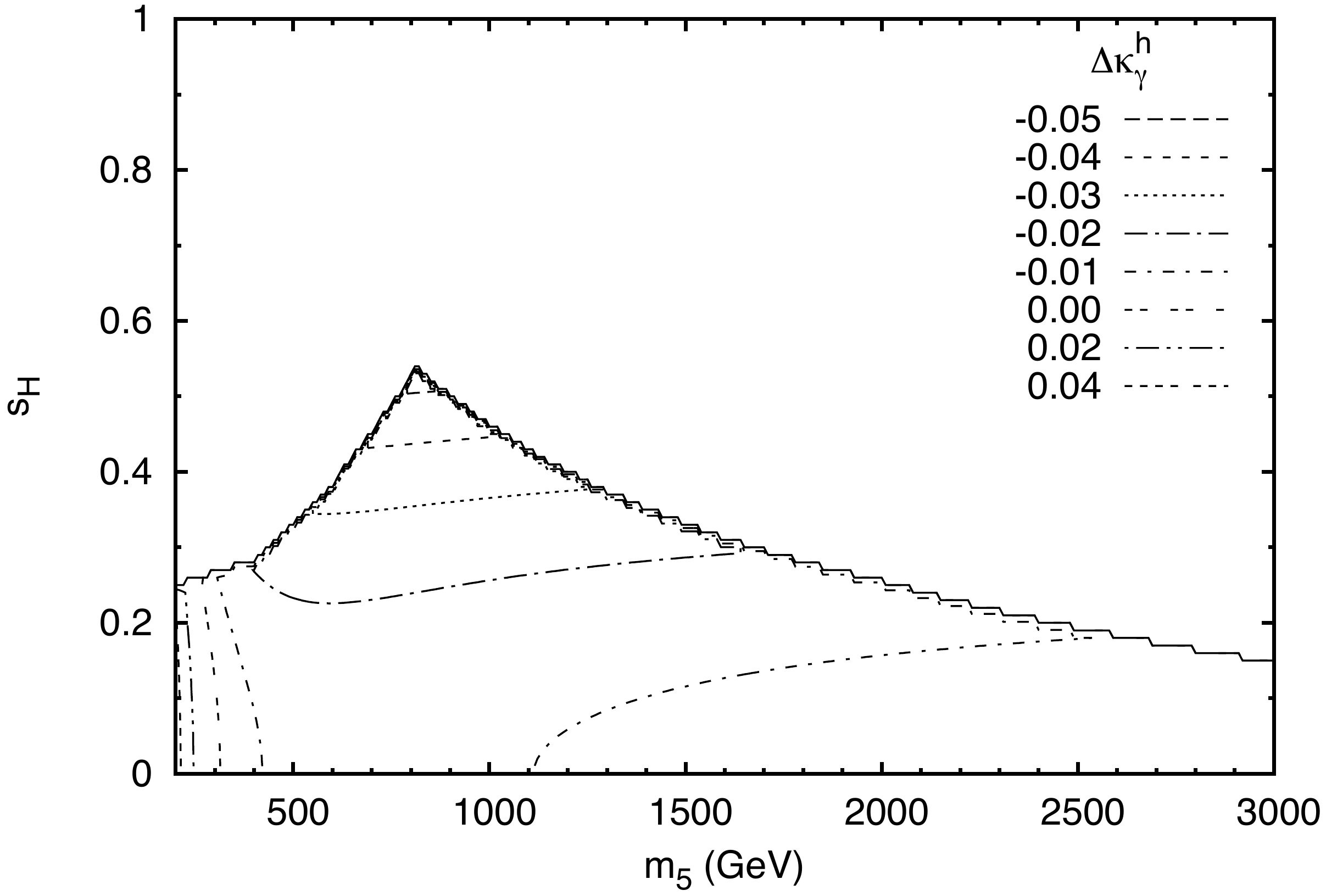}}
\caption{Left: Contours of $\kappa_\gamma^h$ in the H5plane benchmark.
         The value of $\kappa_\gamma^h$ ranges from $0.987$ to $1.24$.
Right: Contours of $\Delta\kappa_\gamma^h$ in the H5plane benchmark.
         The value of $\Delta\kappa_\gamma^h$ ranges from $-0.054$ to $0.052$.}
\label{fig:KGAML_data}
\end{figure}

The coupling of $h$ to photon pairs is affected by the modifications of these tree-level couplings, as well as by contributions from loop diagrams involving $H_3^{\pm}$, $H_5^{\pm}$, and $H_5^{\pm\pm}$.  Defining $\kappa^h_{\gamma}$ in the usual way as~\cite{LHCHiggsCrossSectionWorkingGroup:2012nn}\footnote{In {\tt GMCALC 1.2.1} the computation of the fermion loop contribution to Higgs decays to two photons includes only the top quark loop.}
\begin{equation}
	\kappa^h_{\gamma} = \sqrt{\frac{\Gamma(h \to \gamma\gamma)}{\Gamma(h_{\rm SM} \to \gamma\gamma)}},
\end{equation}
we plot this coupling in the H5plane benchmark in the left panel of Fig.~\ref{fig:KGAML_data}.  The coupling of $h$ to photons $\kappa^h_{\gamma}$ varies between 0.99 and 1.24, reaching its largest values when $s_H$ is large.  To isolate the effect of the loop diagrams involving $H_3^{\pm}$, $H_5^{\pm}$, and $H_5^{\pm\pm}$, in the right panel of Fig.~\ref{fig:KGAML_data} we plot $\Delta \kappa^h_{\gamma}$, which is defined as the contribution to $\kappa^h_{\gamma}$ made by the scalar loops, i.e.,
\begin{equation}
	\Delta \kappa^h_{\gamma} = \kappa^h_{\gamma}({\rm full}) - \kappa^h_{\gamma}(t~{\rm and}~W~{\rm loops~only}).
\end{equation}
$\Delta \kappa_{\gamma}^h$ varies between $\pm0.05$ in the H5plane benchmark.  It is positive only for $m_5$ below 300~GeV, where it contributes to a slight enhancement of $\kappa_{\gamma}^h$ to values up to 1.05.  It reaches its most negative value at large $s_H$, where it limits the enhancement of $\kappa_{\gamma}^h$ through destructive interference with the dominant $W$ loop contribution.

We also examine the total width of $h$ in the H5plane benchmark.  We define the scaling factor $\kappa_h$ as~\cite{LHCHiggsCrossSectionWorkingGroup:2012nn}
\begin{equation}
	\kappa_h = \sqrt{\frac{\Gamma_{\rm tot}(h)}{\Gamma_{\rm tot}(h_{\rm SM})}},
\end{equation}
and calculate it using the formula
\begin{equation}
  \kappa_h^2 = \frac{(\kappa^{h}_f)^2 \left(B^{SM}_{h \to b \bar{b}} + B^{SM}_{h \to \tau^+ \tau^-}
                            + B^{SM}_{h \to c \bar{c}} + B^{SM}_{h \to g g}\right)
                              + (\kappa^{h}_V)^2 \left(B^{SM}_{h \to W^+ W^-} + B^{SM}_{h \to ZZ}\right)
                                + (\kappa^{h}_{\gamma})^2 B^{SM}_{h \to \gamma \gamma}
                                  + (\kappa^{h}_{Z\gamma})^2 B^{SM}_{h \to \gamma Z}}
                        {B^{SM}_{h \to b \bar{b}} + B^{SM}_{h \to \tau^+ \tau^-}
                            + B^{SM}_{h \to c \bar{c}} + B^{SM}_{h \to g g}
                              + B^{SM}_{h \to W^+ W^-} + B^{SM}_{h \to ZZ}
                                + B^{SM}_{h \to \gamma \gamma} + B^{SM}_{h \to \gamma Z}} .
\end{equation}
The values for the SM Higgs branching ratios $B^{SM}_{h \to X}$ were taken from Tables 174--178 of Ref.~\cite{deFlorian:2016spz} for a SM Higgs mass of 125.09~GeV and are reproduced in Table~\ref{tab:BRs}.  We use this more precise value of the SM Higgs boson mass in this calculation because the LHC Higgs coupling measurements in Ref.~\cite{Aad:2016} have been extracted for this mass value.

\begin{table}
  \begin{tabular}{c c}
   \hline \hline
   branching ratio                   & value \\
   \hline
   $B^{SM}_{h \to b \bar{b}}$     & $5.809 \times 10^{-1}$ \\
   $B^{SM}_{h \to \tau^+ \tau^-}$ & $6.256 \times 10^{-2}$ \\
   $B^{SM}_{h \to c \bar{c}}$     & $2.884 \times 10^{-2}$ \\
   $B^{SM}_{h \to g g}$           & $8.180 \times 10^{-2}$ \\
   $B^{SM}_{h \to W^+ W^-}$       & $2.152 \times 10^{-1}$ \\
   $B^{SM}_{h \to ZZ}$            & $2.641 \times 10^{-2}$ \\
   $B^{SM}_{h \to \gamma \gamma}$ & $2.270 \times 10^{-3}$ \\
   $B^{SM}_{h \to \gamma Z}$      & $1.541 \times 10^{-3}$ \\
   \hline \hline
  \end{tabular}
\caption{Branching ratios of the SM Higgs boson with mass 125.09~GeV, from Ref.~\cite{deFlorian:2016spz}, used in the calculation of $\kappa_h$.}
\label{tab:BRs}
\end{table}

\begin{figure}
\resizebox{0.5\textwidth}{!}{\includegraphics{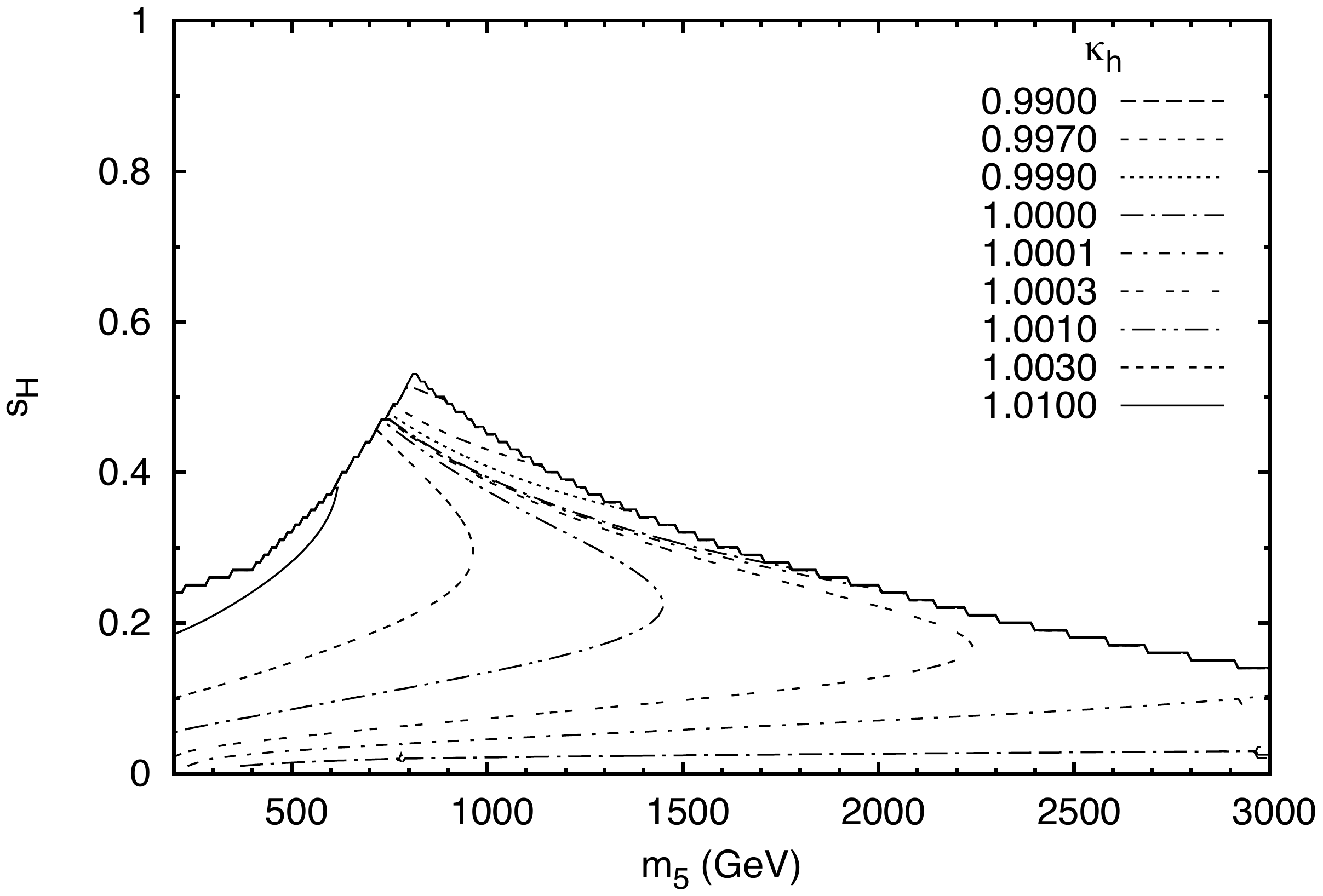}}%
\resizebox{0.5\textwidth}{!}{\includegraphics{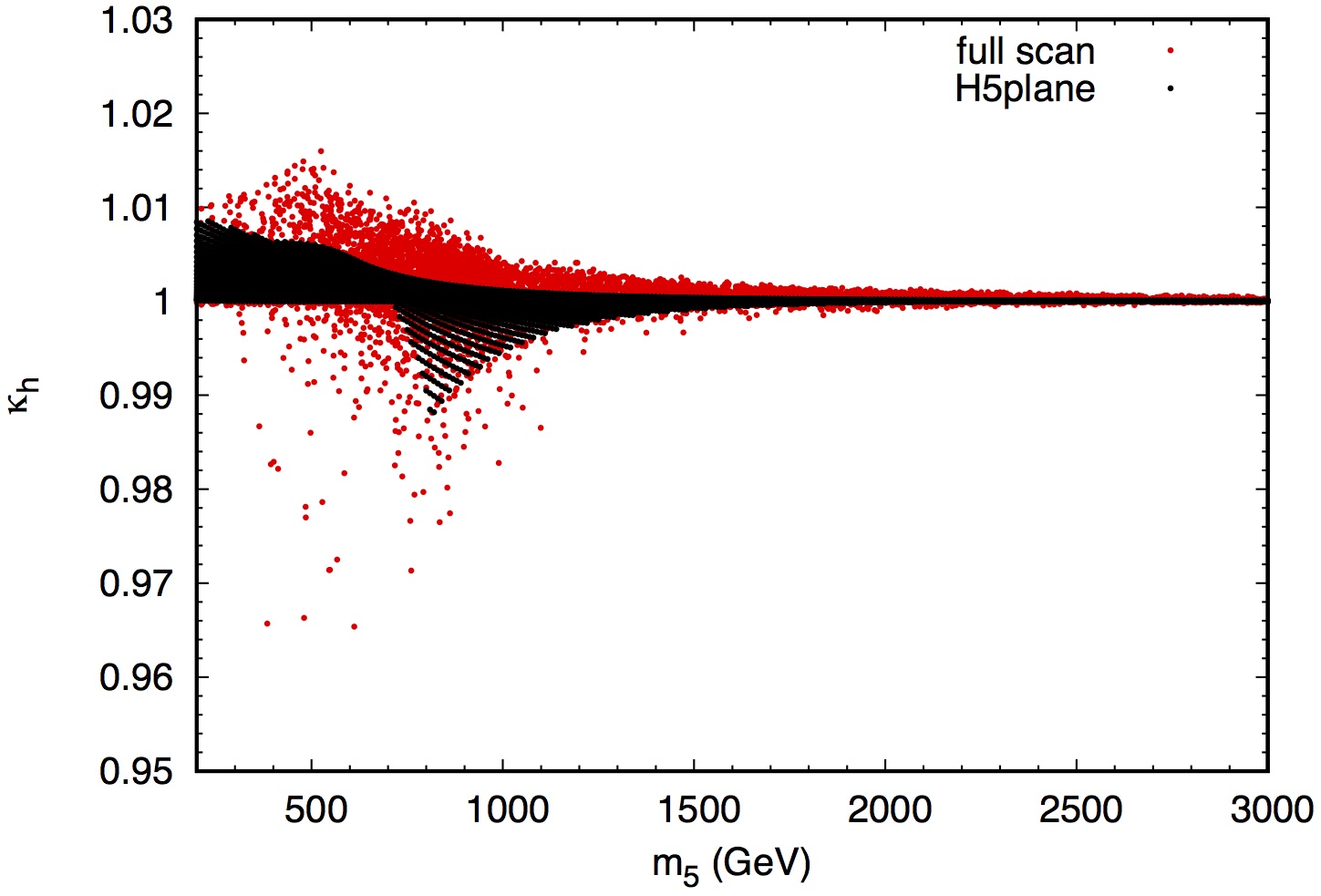}}
\caption{Left: Contours of $\kappa_h$ in the H5plane benchmark.  The value of $\kappa_h$ ranges from $0.985$ to $1.017$.
Right: $\kappa_h$ as a function of $m_5$ in the H5plane benchmark (black points) and in a full scan of the GM model parameter space (red points).  Indirect constraints from $b \to s \gamma$ and the $S$ parameter~\cite{Hartling:2014aga} and direct constraints from a CMS search for $H_5^{\pm\pm} \to W^{\pm}W^{\pm}$ in vector boson fusion~\cite{Khachatryan:2014sta} have been applied.}
\label{fig:kappa_hl_data}
\end{figure}

We plot $\kappa_h$ in the H5plane benchmark in the left panel of Fig.~\ref{fig:kappa_hl_data}.  $\kappa_h$ remains very close to one over the entire benchmark, varying between $0.985$ and $1.017$, which is surprising considering that the tree-level couplings of $h$ to vector bosons are modified by as much as 21\% and those of $h$ to fermions by as much as 10\% compared to the SM Higgs couplings.  The very SM-like values of the $h$ total width are due to an accidental cancellation between an enhancement of the $h$ partial width to vector bosons and a suppression of its partial width to fermions.  This cancellation also occurs, though less severely, in a full scan of the GM model, as shown by the red points in the right panel of Fig.~\ref{fig:kappa_hl_data}.  $\kappa_h$ is slightly greater than one in most of the H5plane benchmark, falling below one in a small sliver at high $s_H$ and $m_5$ between 700 and 1800~GeV, and in a thin band for $s_H < 0.04$.

\begin{figure}
\resizebox{\textwidth}{!}{\includegraphics{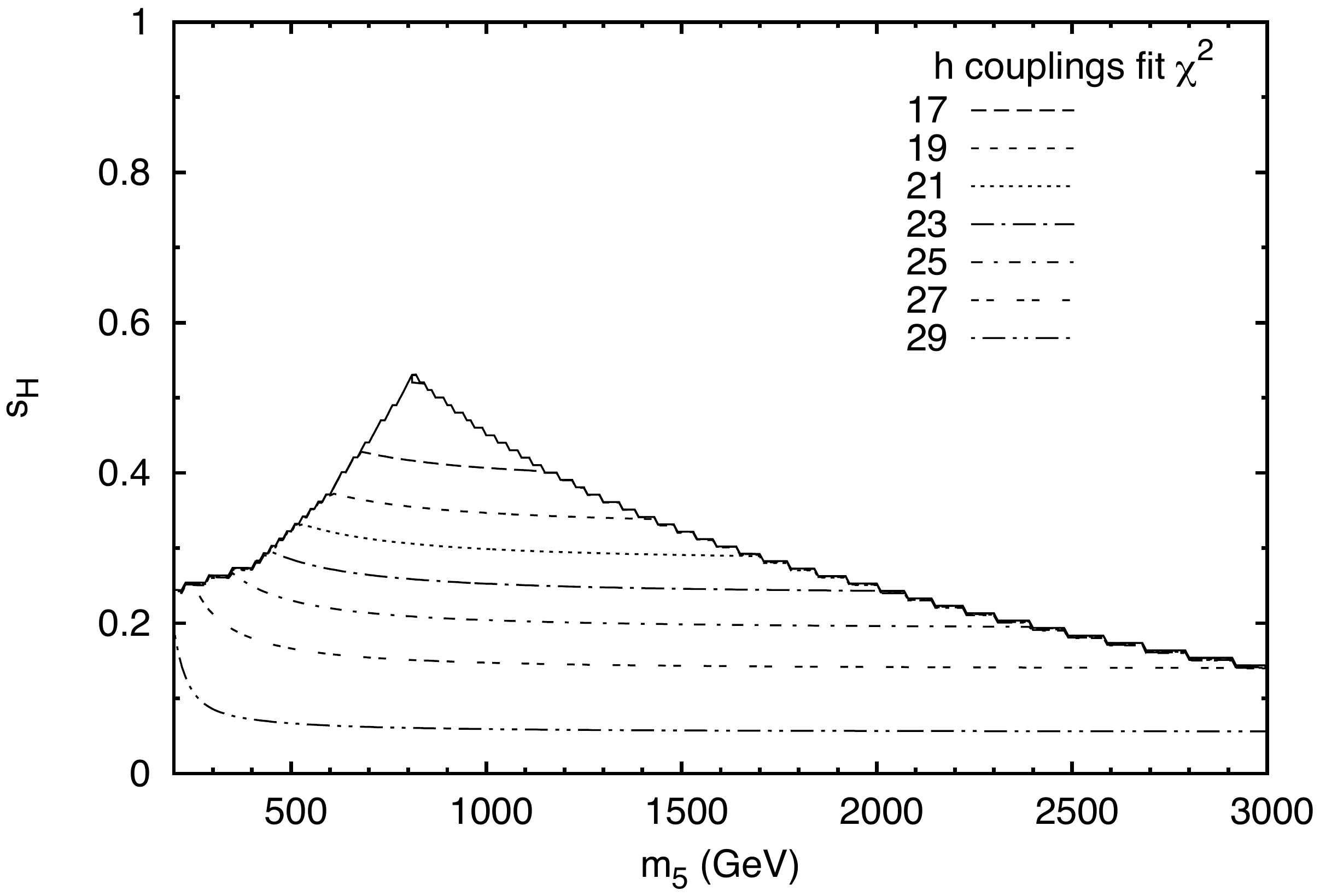}}
\caption{Contours of the $\chi^2$ value for a fit of the $h$ cross sections and branching ratios in the H5plane benchmark to LHC Higgs boson measurements from Ref.~\cite{Aad:2016}.  The $\chi^2$ ranges from $16.2$ to $29.9$.  Compare the $\chi^2$ of 29.4 for the SM Higgs boson.}
\label{fig:chi2_data}
\end{figure}

In order to evaluate the consistency of the H5plane benchmark with LHC measurements of the couplings of the 125~GeV Higgs boson, we compute a $\chi^2$ using the combined ATLAS and CMS Higgs production and decay measurements in Ref.~\cite{Aad:2016} from data collected at LHC centre-of-mass energies of 7 and 8~TeV.  We use the observables and the corresponding correlation matrix $\rho$ summarized in Table~9 and Fig.~28, respectively, of Ref.~\cite{Aad:2016}.  The $\chi^2$ is defined according to
\begin{equation}
 \chi^2 = (\vec{x}-\vec{\mu})^T V^{-1} (\vec{x}-\vec{\mu}), \quad V_{ij} = \rho_{ij} \sigma_i \sigma_j,
\end{equation}
where $\vec{x}$ is the vector of observed values, $\vec{\mu}$ is the vector of theoretical values at a particular point in the H5plane benchmark, and $\vec{\sigma}$
is the vector of the combined theoretical and experimental uncertainties.  Where the experimental uncertainties in Table~9 of Ref.~\cite{Aad:2016} are asymmetric, we symmetrize them by averaging the upper and lower uncertainty.  We then combine the (symmetrized) experimental uncertainties with the theoretical uncertainties quoted in Table~9 of Ref.~\cite{Aad:2016} in quadrature.
The results are shown in Fig.~\ref{fig:chi2_data}.  The $\chi^2$ in the H5plane benchmark of the GM model ranges from a maximum of 29.9 for $s_H$ near zero to a minimum of 16.2 for $s_H$ around 0.5 and $m_5$ around 800--1000~GeV.  For comparison, the $\chi^2$ for the SM Higgs, computed in the same way, is 29.4.  The lower $\chi^2$ values in the GM model reflect a pull in the data towards slightly lower $\kappa_f^h$ and higher $\kappa_V^h$ values.  In particular, we observe that the entire H5plane benchmark is currently consistent with LHC Higgs coupling data.

\subsection{Couplings and decays of $H$}

\begin{figure}
\resizebox{0.5\textwidth}{!}{\includegraphics{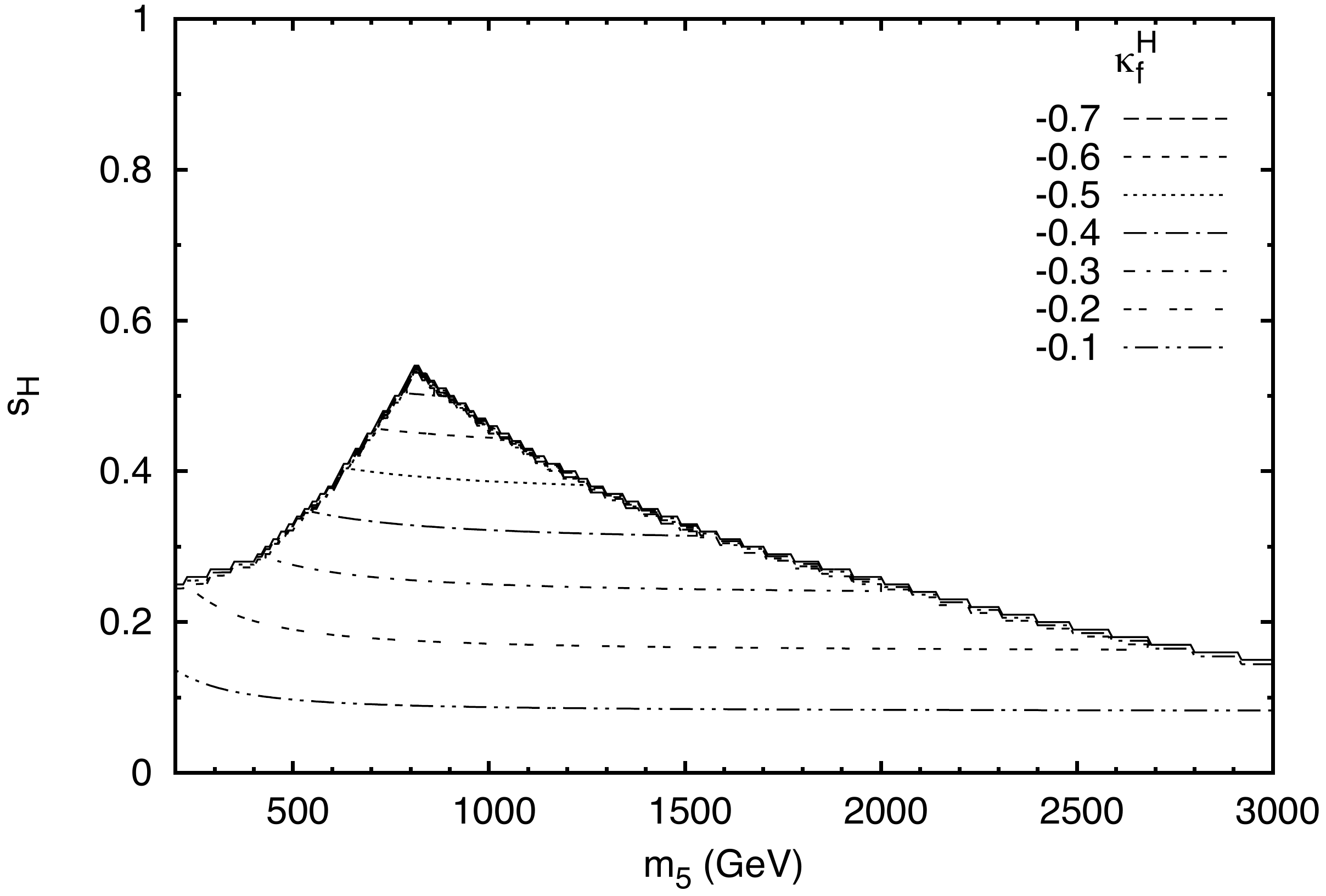}}%
\resizebox{0.5\textwidth}{!}{\includegraphics{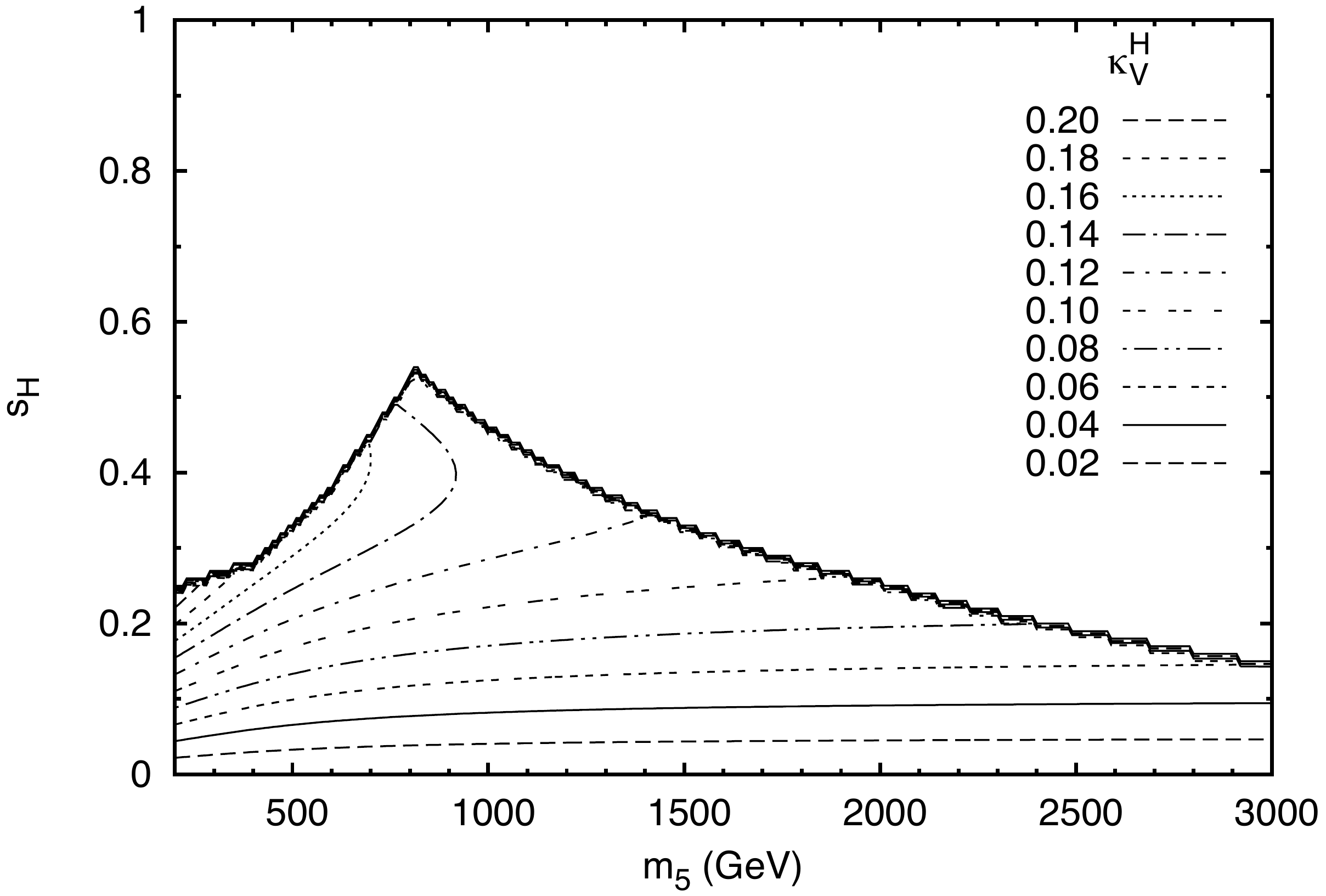}}
\caption{Contours of $\kappa_f^H$ (left) and $\kappa_V^H$ (right) in the H5plane benchmark.
         $\kappa_f^H$ ranges from zero to $-0.76$ and
          $\kappa_V^H$ ranges from zero to $0.22$.}
\label{fig:KVH_data}
\end{figure}

We now examine the couplings and decays of the heavier custodial-singlet Higgs boson $H$.  The tree-level couplings of $H$ in the GM model are given in terms of the underlying parameters by
\begin{equation}
	\kappa_f^H = \frac{s_{\alpha}}{c_H}, \qquad \qquad 
	\kappa_V^H = s_{\alpha} c_H + \sqrt{\frac{8}{3}} c_{\alpha} s_H,
\end{equation}
where the $\kappa$ factors are again defined as the ratio of the $H$ coupling in the GM model to the corresponding coupling of the SM Higgs boson.  In Fig.~\ref{fig:KVH_data} we plot $\kappa_f^H$ (left panel) and $\kappa_V^H$ (right panel) in the H5plane benchmark.  These couplings are interesting mainly because they control the production of $H$ via gluon fusion and vector boson fusion, respectively.  The coupling of $H$ to fermions is largest in magnitude at large $s_H$, reaching $-0.76$ times the corresponding SM Higgs coupling.  The coupling of $H$ to vector boson pairs is largest at low $m_5 \sim 200$--300~GeV and large $s_H$, reaching 0.22 times the corresponding SM Higgs coupling strength.  Squaring these, the cross sections for $H$ production by gluon fusion and vector boson fusion reach at most 0.58 and 0.048 times the corresponding SM Higgs cross sections for a Higgs boson of the same mass as $H$, respectively.

\begin{figure}
\resizebox{0.5\textwidth}{!}{\includegraphics{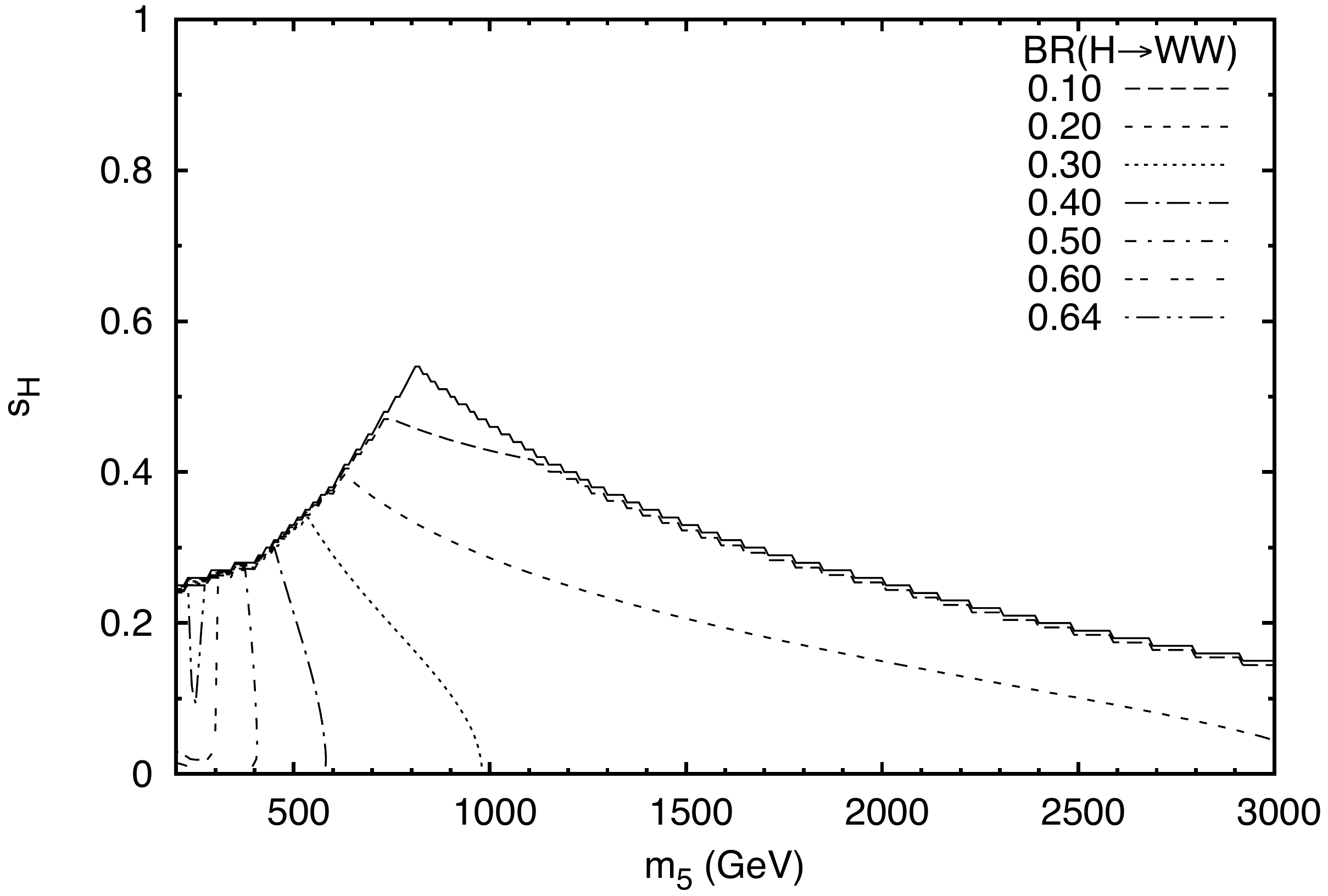}}%
\resizebox{0.5\textwidth}{!}{\includegraphics{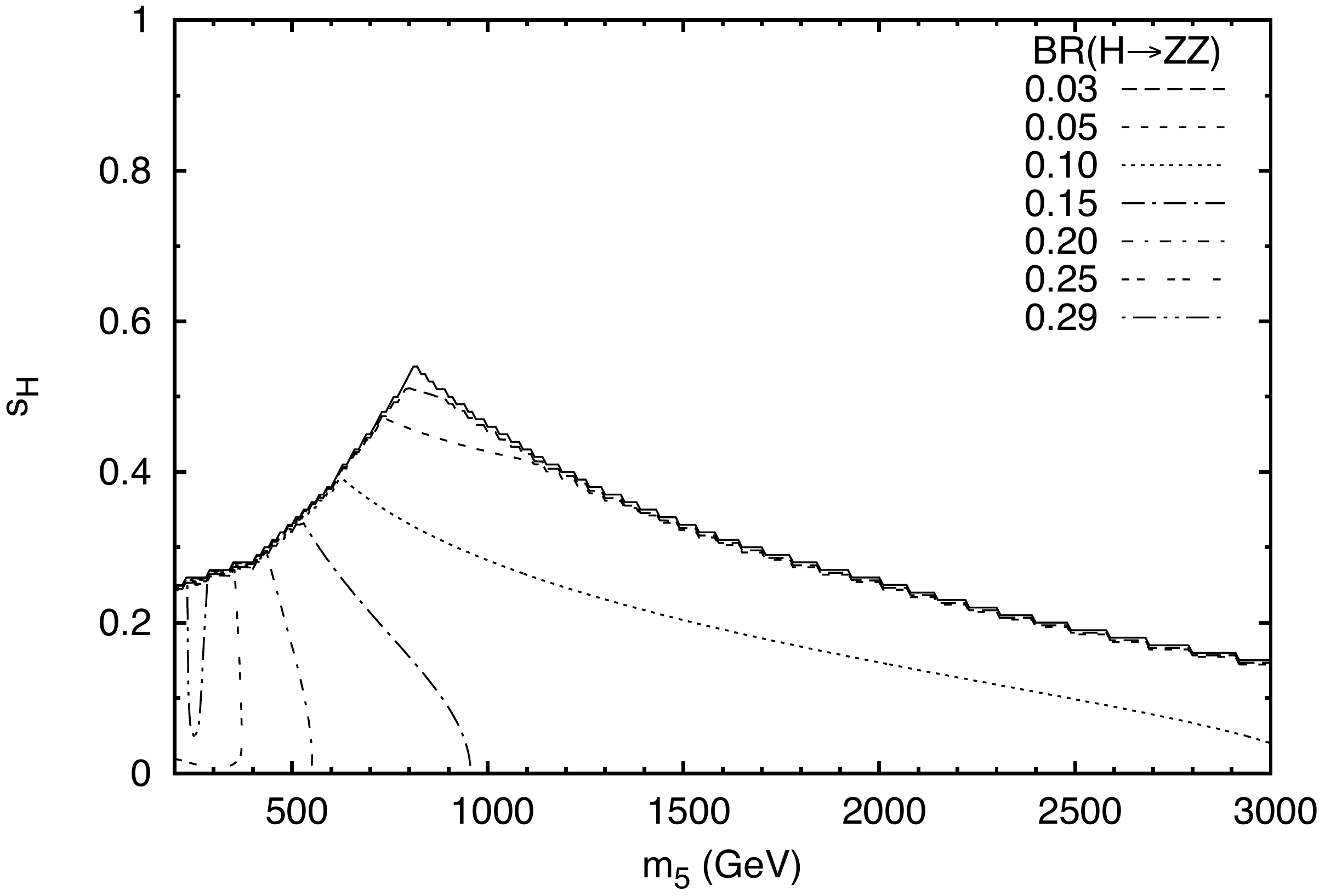}}
\caption{Contours of BR($H \rightarrow W^+ W^-$) (left) and BR($H \rightarrow ZZ$) (right) in the H5plane benchmark.
         BR($H \rightarrow W^+ W^-$) ranges from $0.05$ to $0.65$ and
         BR($H \rightarrow ZZ$) ranges from $0.02$ to $0.30$.}
\label{fig:HHBRW_data}
\end{figure}

\begin{figure}
\resizebox{0.5\textwidth}{!}{\includegraphics{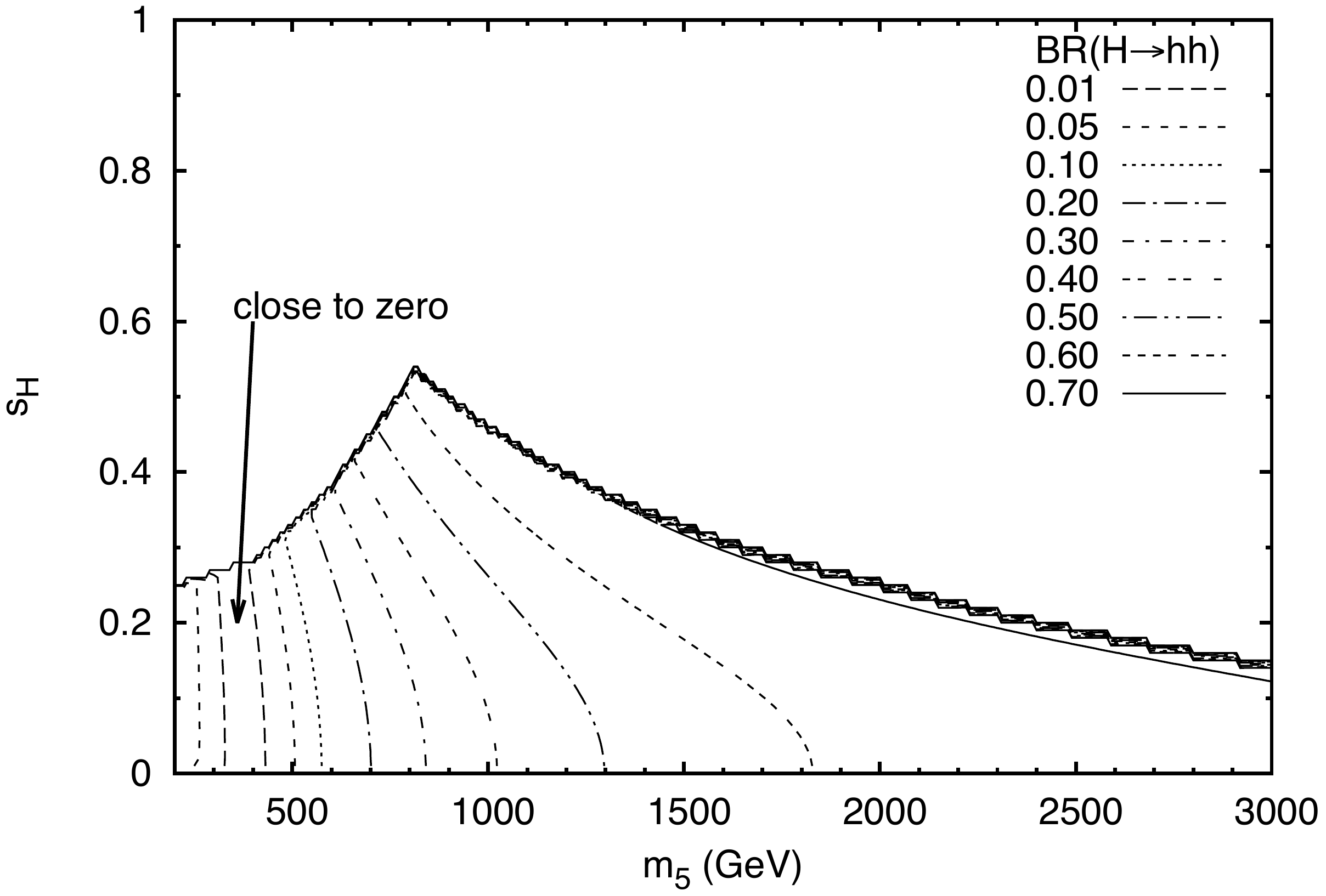}}%
\resizebox{0.5\textwidth}{!}{\includegraphics{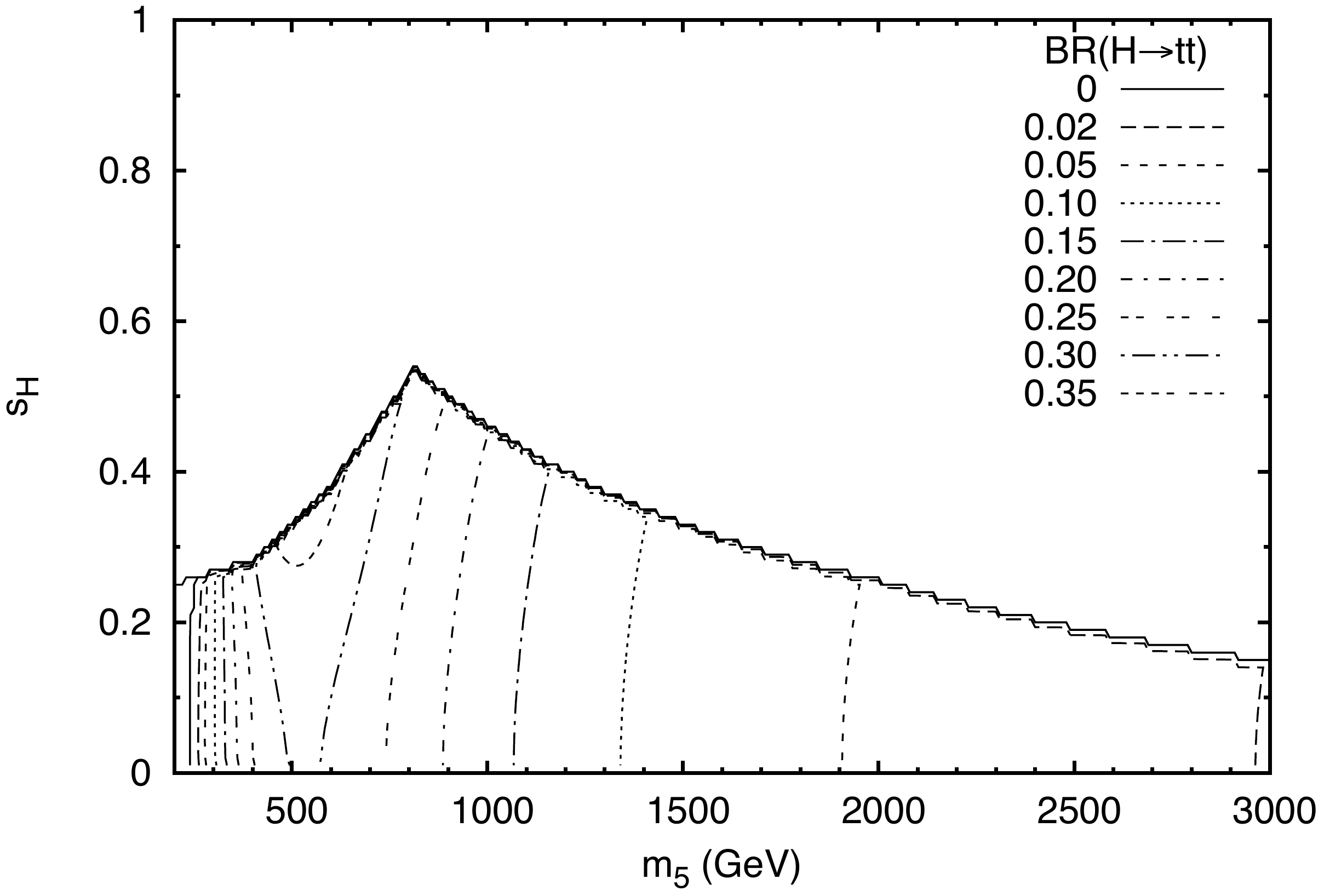}}
\caption{Contours of BR($H \rightarrow hh$) (left) and BR($H \rightarrow t\bar{t}$) (right) in the H5plane benchmark.
         BR($H \rightarrow hh$) ranges from zero to $0.71$ and
         BR($H \rightarrow t\bar{t}$) ranges from zero to $0.37$.
         BR($H \rightarrow t\bar{t}$) drops abruptly to zero when $m_H < 2m_t$ because off-shell decays to $t \bar t$ are not calculated in {\tt GMCALC 1.2.1}.}
\label{fig:HHBRHL_data}
\end{figure}

In Figs.~\ref{fig:HHBRW_data} and \ref{fig:HHBRHL_data} we plot the branching ratios of $H$ to $W^+W^-$, $ZZ$, $hh$, and $t \bar t$.  These are the dominant decays of $H$ over the entire H5plane benchmark.  The branching ratios of $H$ to $W^+W^-$ and $ZZ$ dominate for $m_5$ below 600~GeV, with branching ratios above 40\% and 20\%, respectively.  These decays reach maximum branching ratios of 65\% and 30\%, respectively, for low $m_5 \sim 200$--300~GeV.  The branching ratio of $H$ to $W^+W^-$ ($ZZ$) remains above 20\% (10\%) over most of the benchmark plane, out to the highest $m_5$ values.

The branching ratio of $H$ to $hh$ dominates at high masses, reaching 50\% for $m_5 \sim 1000$~GeV and a maximum of 71\% for the highest $s_H$ values at large $m_5 > 1500$~GeV.  The branching ratio of $H$ to $t \bar t$ reaches a maximum of 37\% for $m_5 \sim 500$--600~GeV and high $s_H$, but falls below 10\% for $m_5 \gtrsim 1400$~GeV.  Note that, because $m_H > m_5$ in the H5plane benchmark, the kinematic threshold for $H \to t \bar t$ at $m_H = 2m_t$ occurs when $m_5 \simeq 250$~GeV.

\subsection{$H$--$H_5$ mass splitting}

\begin{figure}
\resizebox{0.5\textwidth}{!}{\includegraphics{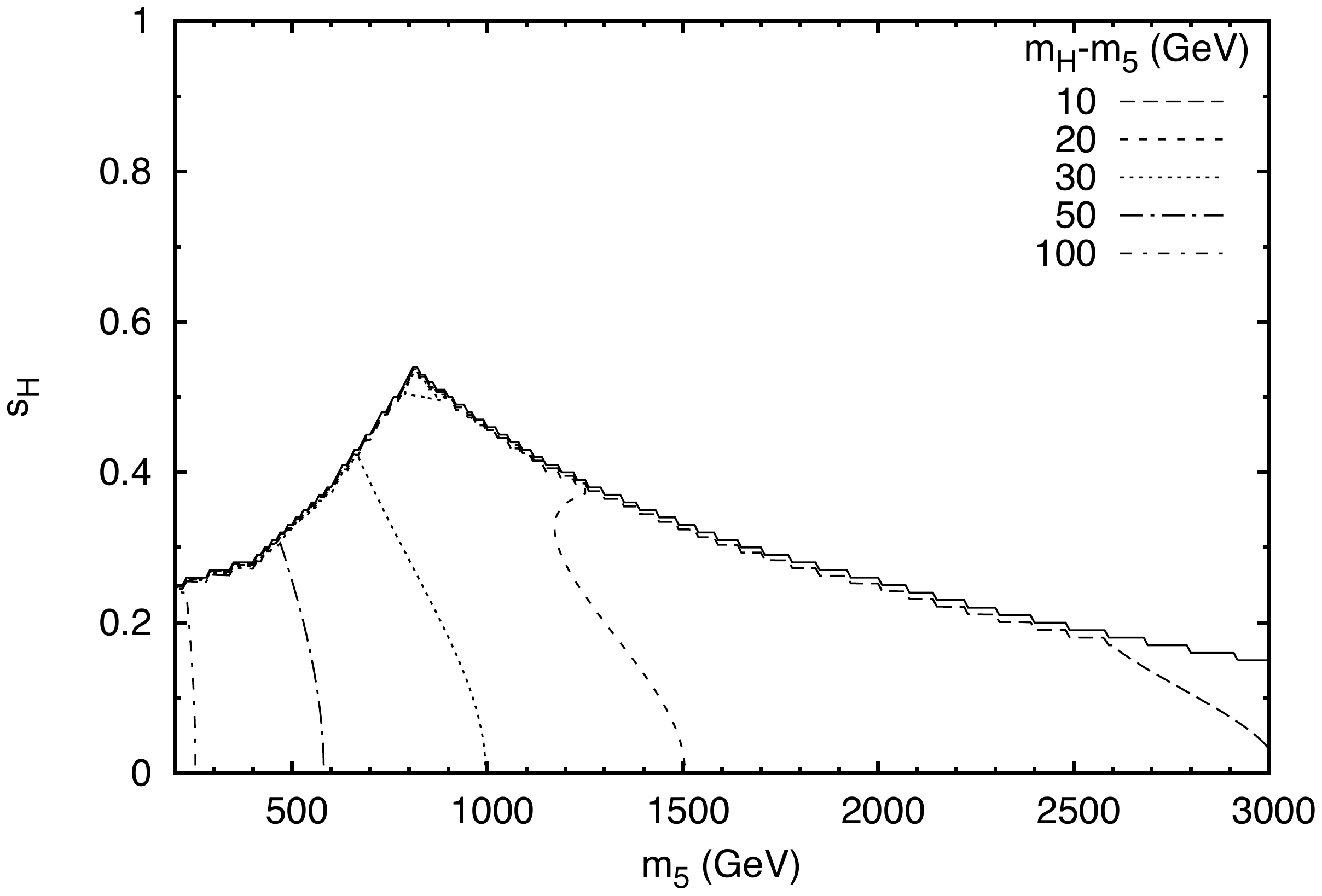}}%
\resizebox{0.5\textwidth}{!}{\includegraphics{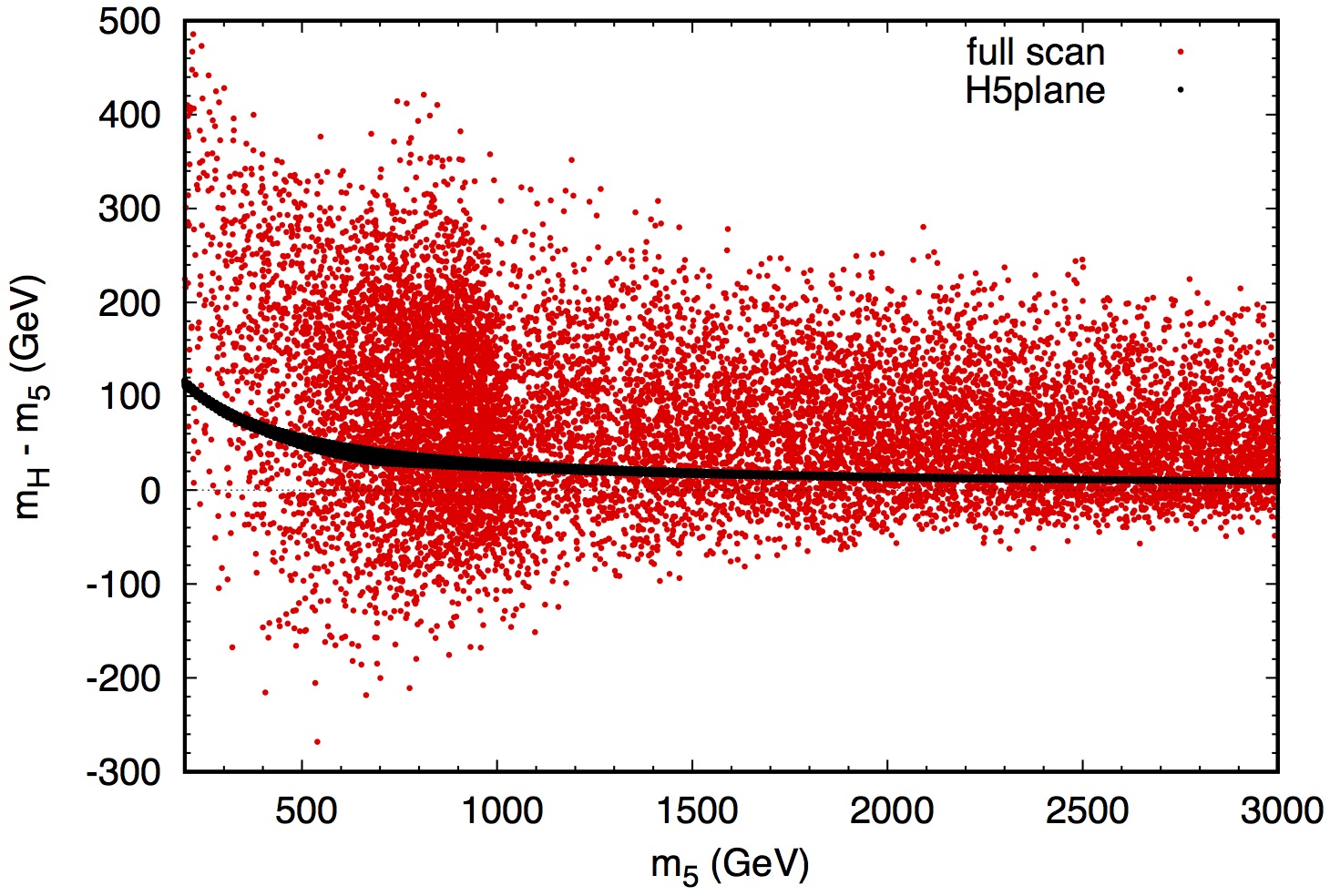}}
\caption{Left: Contours of $m_H-m_5$ in the H5plane benchmark.
         $m_H-m_5$ ranges from $8.9$~GeV to $120$~GeV.
Right: Mass difference $m_H - m_5$ as a function of $m_5$ in the H5plane benchmark (black points) and in a full scan of the GM model parameter space (red points).  Indirect constraints from $b \to s \gamma$ and the $S$ parameter~\cite{Hartling:2014aga} and direct constraints from a CMS search for $H_5^{\pm\pm} \to W^{\pm}W^{\pm}$ in vector boson fusion~\cite{Khachatryan:2014sta} have been applied.}
\label{fig:mH-m5_fullscan}
\end{figure}

Decays of $H_5^+$ to $H W^+$ and of $H_5^0$ to $HZ$ or $HH$ are forbidden by custodial symmetry.
Therefore our interest in the mass splitting between $H$ and $H_5^0$ is due to the fact that both of these states can be produced in vector boson fusion with decays to $W^+W^-$ and $ZZ$, which opens the possibility of interference between their lineshapes if the resonances are close enough together.  In the left panel of Fig.~\ref{fig:mH-m5_fullscan} we show the mass splitting $m_H - m_5$ in the H5plane benchmark.  The splitting varies from 120~GeV at $m_5 = 200$~GeV to about 9~GeV at $m_5 = 3000$~GeV.  In the right panel of Fig.~\ref{fig:mH-m5_fullscan} we plot $m_H - m_5$ as a function of $m_5$ scanning over all the other free parameters in the H5plane benchmark (black points) and the full GM model (red points), where we have imposed the indirect constraints from $b \to s \gamma$ and the $S$ parameter~\cite{Hartling:2014aga} and direct constraints from the CMS search for $H_5^{\pm\pm} \to W^{\pm}W^{\pm}$ in vector boson fusion~\cite{Khachatryan:2014sta}.  Similarly to the case of $m_3-m_5$, we see that the variation in the mass difference $m_H - m_5$ is much greater in the full model scan than it is in the H5plane benchmark.

\begin{figure}
\resizebox{0.5\textwidth}{!}{\includegraphics{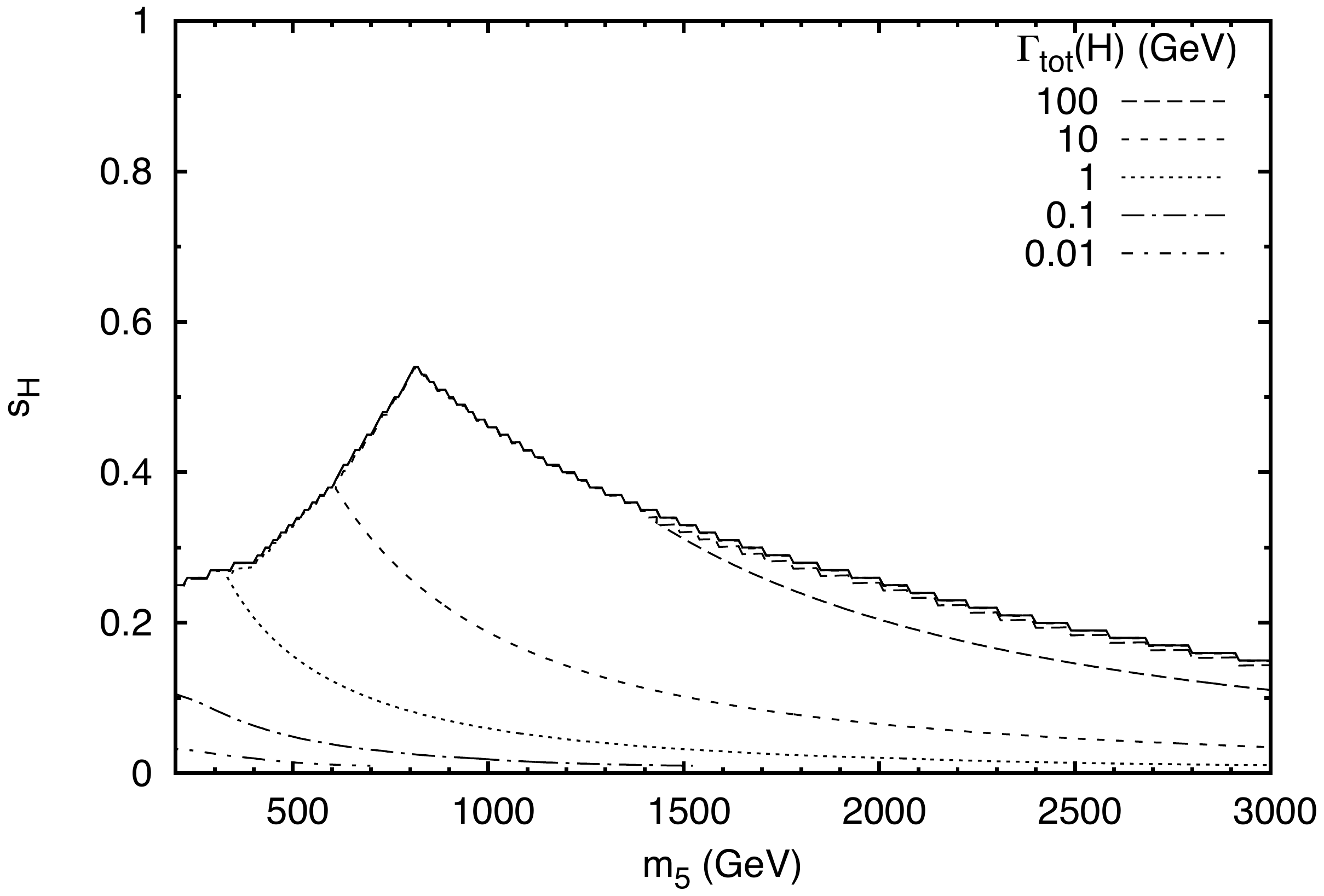}}%
\resizebox{0.5\textwidth}{!}{\includegraphics{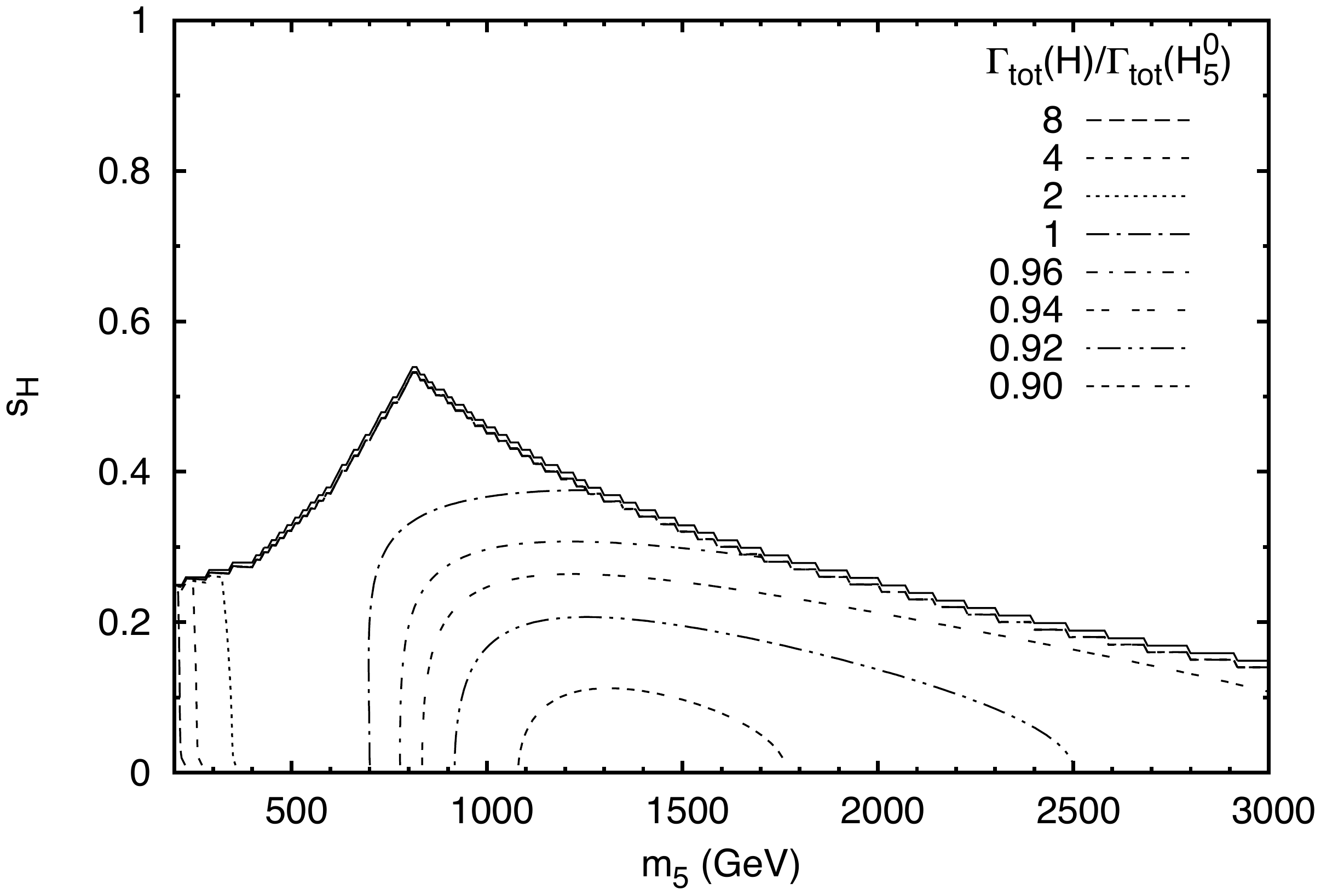}} \\
\resizebox{0.5\textwidth}{!}{\includegraphics{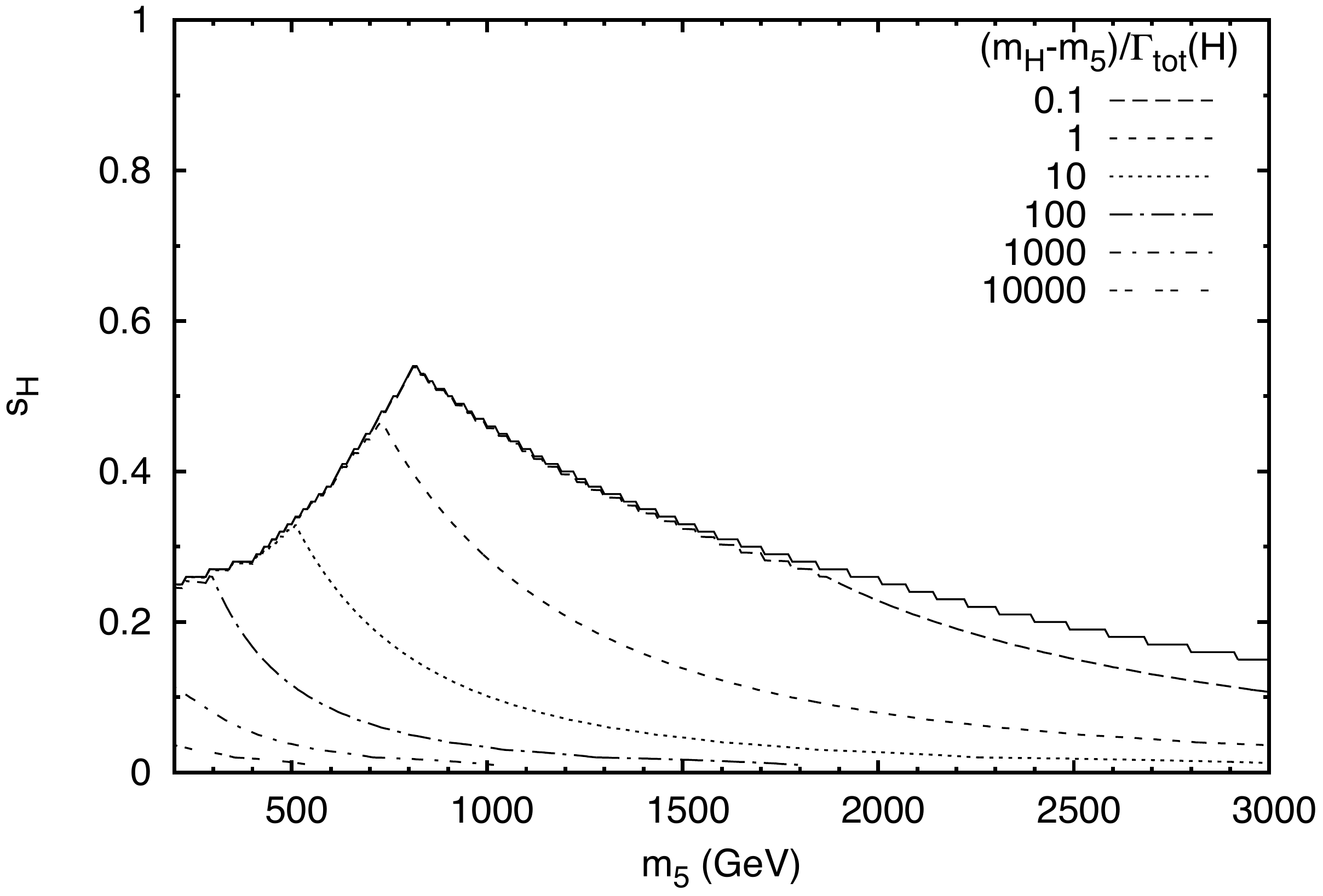}}
\caption{Top left: Contours of the total width of $H$, $\Gamma_{\rm tot}(H)$, in the H5plane benchmark.
         $\Gamma_{\rm tot}(H)$ ranges from $0.0013$~GeV to $170$~GeV.
Top right: Contours of the ratio $\Gamma_{\rm tot}(H)/\Gamma_{\rm tot}(H_5^0)$ in the H5plane benchmark.
         $\Gamma_{\rm tot}(H)/\Gamma_{\rm tot}(H_5^0)$ ranges from  $0.89$ to    $16$.
Bottom: Contours of $(m_H - m_5) / \Gamma_{\rm tot}(H)$ in the H5plane benchmark.
         $(m_H - m_5) / \Gamma_{\rm tot}(H)$ ranges from $0.054$  to    $89000$.}
\label{fig:twHH_data}
\end{figure}

To understand the experimental implications of this mass splitting, we compare it to the intrinsic widths of $H$ and $H_5^0$.  In Fig.~\ref{fig:twHH_data} we first plot the total width of $H$ (top left panel) and the ratio $\Gamma_{\rm tot}(H)/\Gamma_{\rm tot}(H_5^0)$ (top right panel) in the H5plane benchmark.  The total widths of $H$ and $H_5^0$ are very similar for $m_5 \gtrsim 500$ GeV.  For lower masses, the fact that $H$ is significantly heavier than $H_5^0$ allows its width to become more than twice as large as that of $H_5^0$ for $m_5 < 450$~GeV.  Over the entire H5plane benchmark, the width of $H$ is never less than 89\% of the width of $H_5^0$.

Therefore we can quantify the $H$--$H_5^0$ mass splitting by comparing it to the total width of $H$.  We do this in the bottom panel of Fig.~\ref{fig:twHH_data}, in which we plot $(m_H-m_5)/\Gamma_{\rm tot}(H)$ over the H5plane benchmark.  This ratio varies widely over the benchmark.  For low $m_5$ and low $s_H$, $(m_H-m_5)/\Gamma_{\rm tot}(H)$ is large, which means that the $H$ and $H_5^0$ resonances are well separated compared to their intrinsic widths.  However, there is a sizable region of parameter space in which $(m_H-m_5)/\Gamma_{\rm tot}(H) < 1$, which means that the mass splitting is less than the intrinsic width of $H$.  In this region of the H5plane benchmark, the total width of $H_5^0$ is within 10\% of that of $H$.  In this case the two resonances overlap significantly and interfere, so that experimental searches for these two states in vector boson fusion with decays to $W^+ W^-$ or $ZZ$ must be performed taking into account both resonances and their interference.
Interference can be avoided by searching for $H$ produced in gluon fusion, or decaying to $hh$ or $t \bar t$.

\subsection{Decays of $H_3$}

\begin{figure}
\resizebox{0.5\textwidth}{!}{\includegraphics{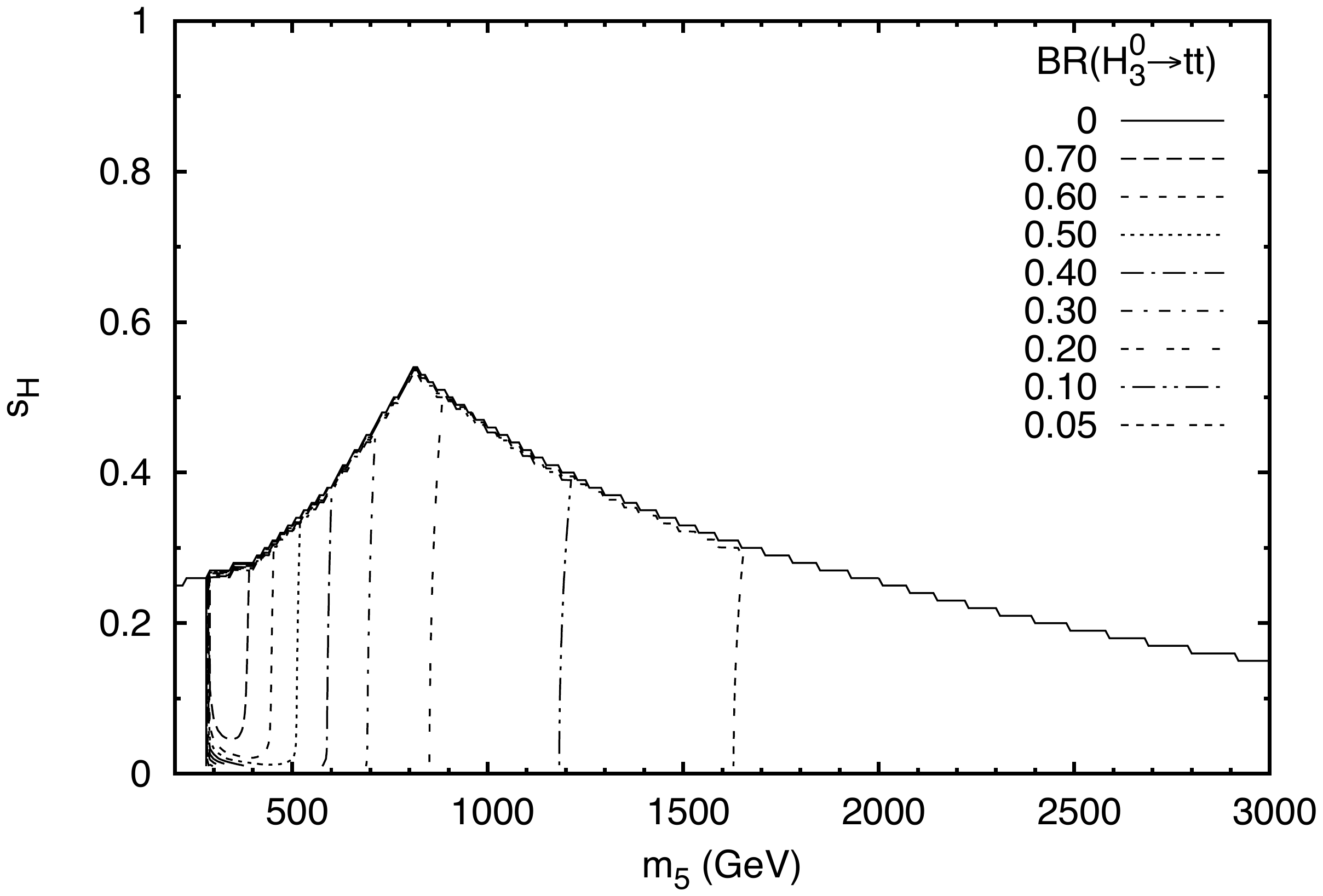}}%
\resizebox{0.5\textwidth}{!}{\includegraphics{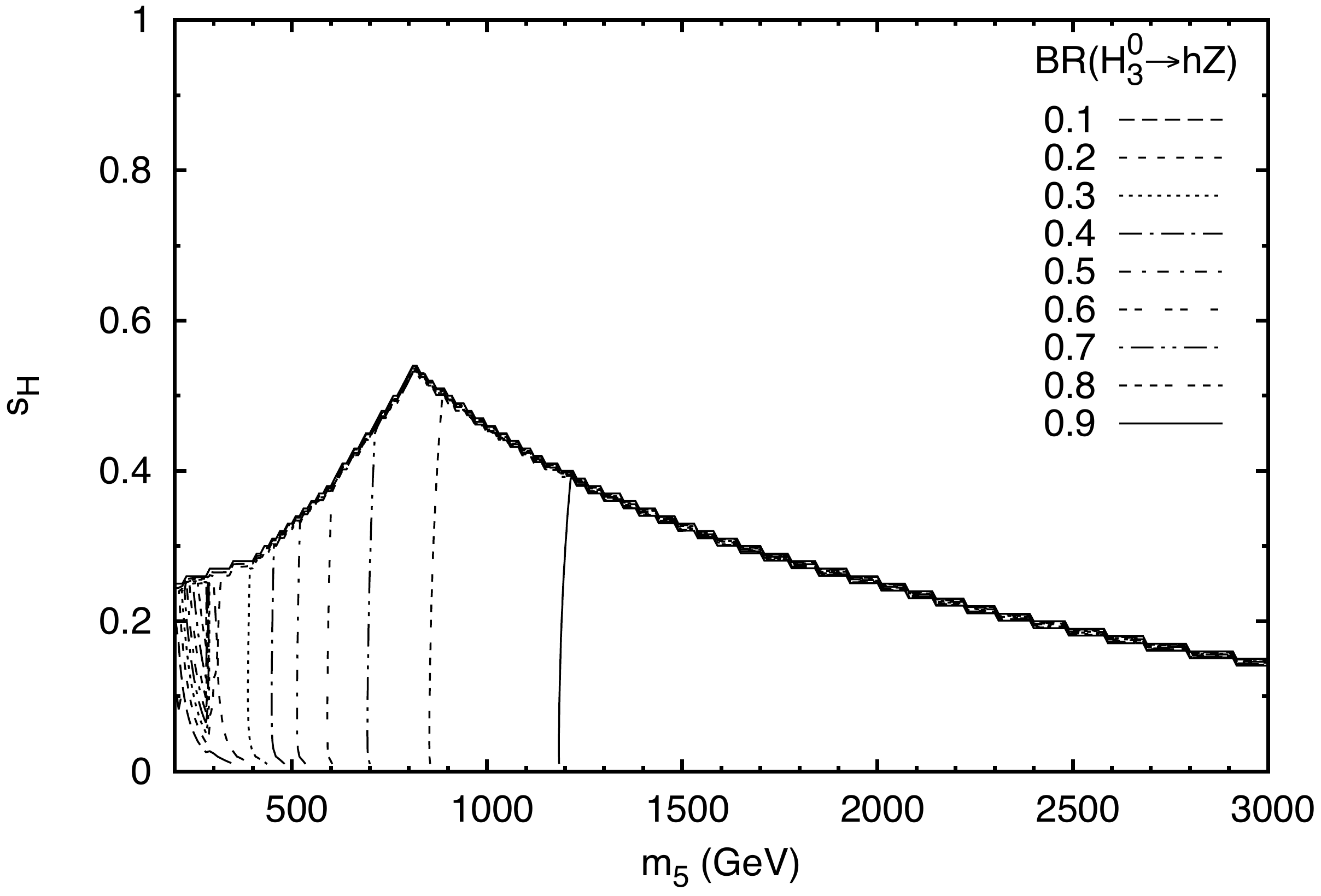}}
\caption{Left: Contours of BR($H_3^0 \rightarrow t\bar{t}$) in the H5plane benchmark.
         BR($H_3^0 \rightarrow t\bar{t}$) ranges from zero  to $0.79$.
         Below the kinematic threshold at $m_3 = 2m_t$, the branching ratio drops to zero because off-shell decays to $t \bar t$ are not calculated in {\tt GMCALC 1.2.1}.
         BR($H_3^0 \to t \bar t$) reaches a maximum of 0.79 and falls to 0.013 at $m_5 = 3000$~GeV.
Right: Contours of BR($H_3^0 \rightarrow h Z$) in the H5plane benchmark.
         BR($H_3^0 \rightarrow h Z$) ranges from $2 \times 10^{-4}$ to $0.987$.
         In the band $m_5 \in (200~\text{GeV}, 300~\text{GeV})$, the branching ratio increases rapidly,
         up to nearly $0.9$ for $m_5=280~\text{GeV}$, before collapsing down to about $0.2$; BR($H_3^0 \rightarrow h Z$) then rises with increasing $m_5$.  The sudden drop in BR($H_3^0 \to h Z$) is due to crossing the kinematic threshold for $H_3^0 \rightarrow t\bar{t}$.}
\label{fig:H3NBRT_data1}
\end{figure}

\begin{figure}
\resizebox{0.5\textwidth}{!}{\includegraphics{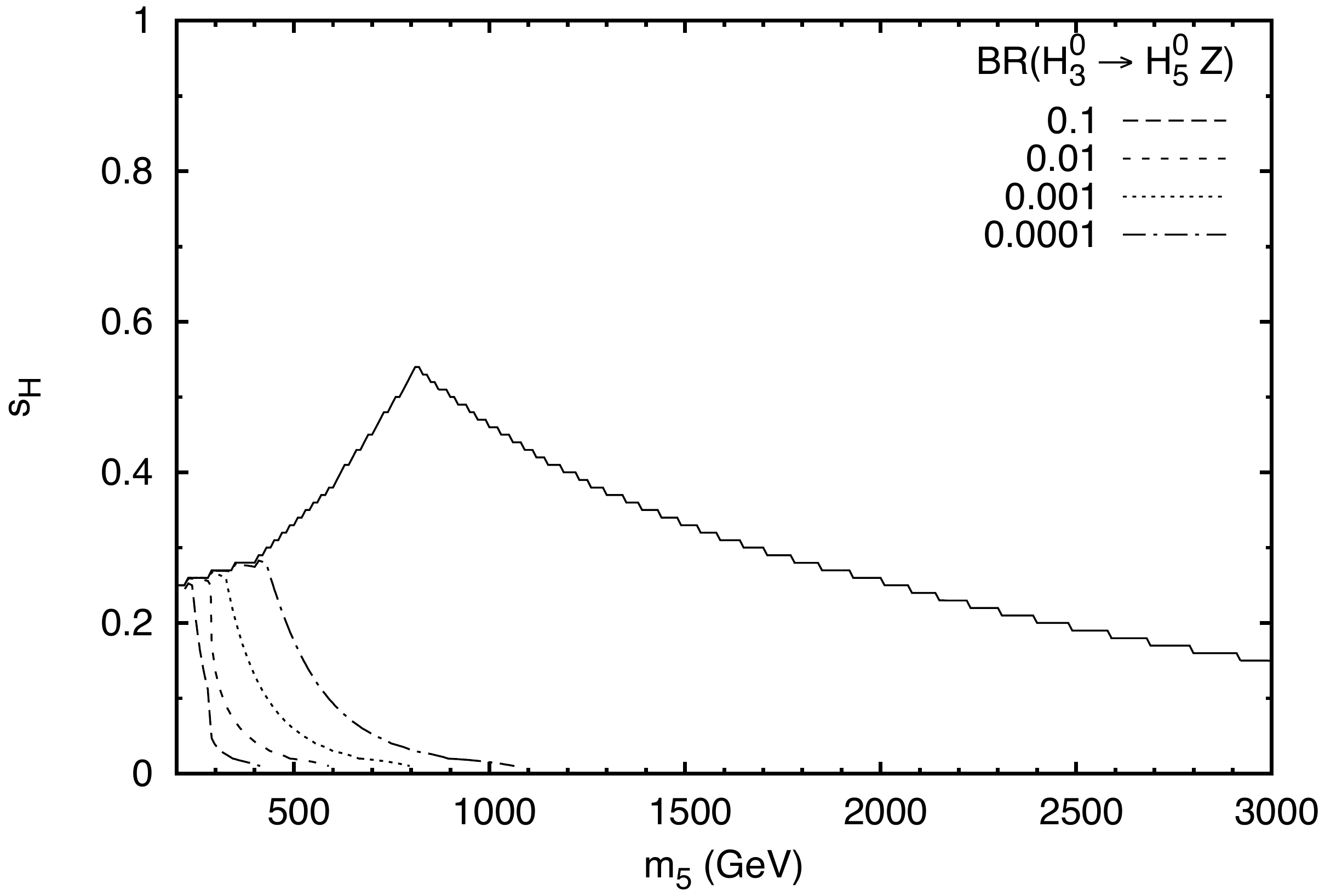}}%
\resizebox{0.5\textwidth}{!}{\includegraphics{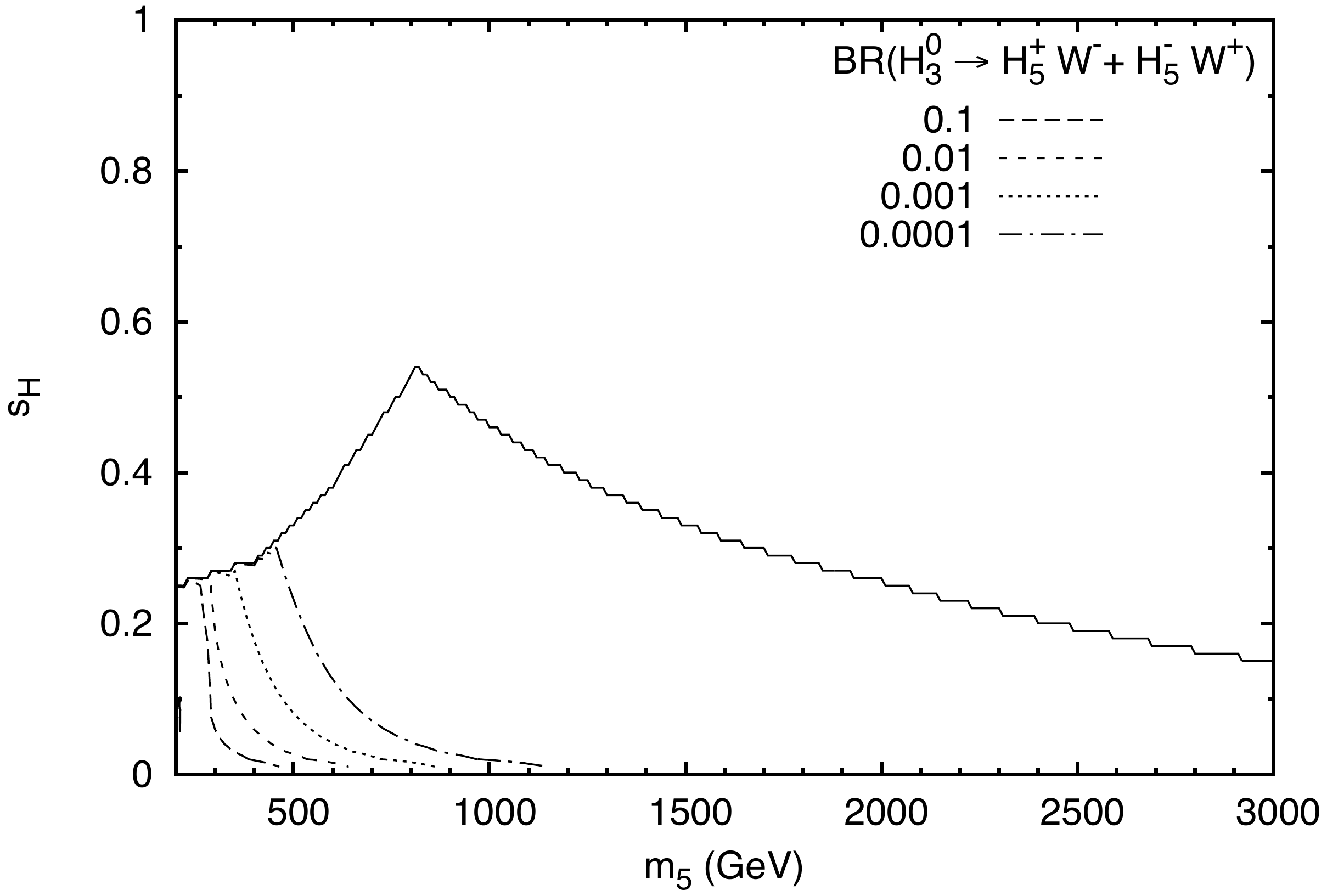}}
\caption{Contours of BR($H_3^0 \rightarrow H_5^0 Z$) (left) and BR($H_3^0 \rightarrow H_5^+ W^- + H_5^- W^+$) (right) in the H5plane benchmark.
See text for further discussion.}
\label{fig:H3NBRZH5N_data}
\end{figure}

The dominant decays of $H_3^0$ in the H5plane benchmark are to $t \bar t$, $hZ$, $H_5^0 Z$, and $H_5^{\pm} W^{\mp}$.  ($H_3^0$ can also decay to two photons; however, BR($H_3^0 \rightarrow \gamma \gamma$) stays below $1.8 \times 10^{-4}$ over the entire H5plane benchmark.)
We plot the branching ratios for these modes in Figs.~\ref{fig:H3NBRT_data1} and \ref{fig:H3NBRZH5N_data}.  The kinematic threshold for $H_3^0 \to t \bar t$ at $m_3 = 2 m_t$ occurs at $m_5$ just below 300~GeV.  Once above this threshold, BR($H_3^0 \to t \bar t$) quickly rises to a maximum of 79\% for $m_5 \sim 300$--400~GeV, and then falls with increasing $m_5$.  The next-largest fermionic decay branching ratio of $H_3^0$ is to $b \bar b$, which is below 1\% over almost all of the H5plane benchmark.  The branching ratio of $H_3^0$ to $h Z$ exhibits complementary behaviour, growing with $m_5$ to become the dominant decay mode ($>50\%$) for $m_5 \gtrsim 500$~GeV and surpassing 90\% branching ratio for $m_5 \gtrsim 1200$~GeV.

The branching ratios of $H_3^0$ to $H_5^0 Z$ and $H_5^{\pm} W^{\mp}$ (we plot the sum of the branching ratios to $H_5^+W^-$ and $H_5^- W^+$) are significant only for very low $m_5$, below the kinematic threshold for the $t \bar t$ decay.  For these low masses, the branching ratios of these modes can be quite large, reaching respective values of 85\% and 82\% in our calculation, in slightly different areas of parameter space.  However, these numbers should be treated with caution because the implementation in {\tt GMCALC 1.2.1} of scalar decays to scalar plus vector at and below the kinematic threshold is still rather primitive.  At $m_5 = 200$~GeV, the mass splitting between $H_3$ and $H_5$ in the H5plane benchmark is 84~GeV, so that the on-shell decay $H_3^0 \to H_5^{\pm} W^{\mp}$ is barely kinematically allowed, while $H_3^0 \to H_5^0 Z$ is off shell.  As $m_5$ increases, the mass splitting decreases, and $H_3^0 \to H_5^{\pm} W^{\mp}$ goes off shell at $m_5 \simeq 210$~GeV.  Above threshold, {\tt GMCALC 1.2.1} computes these decay widths using the two-body on-shell decay formula, while below threshold the computation takes into account the offshellness of the vector boson only.  This is a reasonable approximation at $m_5 \sim 200$~GeV where the $H_5$ scalars are very narrow; however, the transition from the on-shell to off-shell decay widths is not smooth.  The handling of this transition, along with off-shell decays of $H_3^0 \to t \bar t$, should be improved if detailed predictions for the $H_3^0$ branching ratios for $m_5 \lesssim 280$~GeV are needed.  The branching ratios for $H_3^0$ to $H_5^0 Z$ and $H_5^{\pm} W^{\mp}$ fall below 1\% for $m_5 \gtrsim 500$~GeV.

\begin{figure}
\resizebox{0.5\textwidth}{!}{\includegraphics{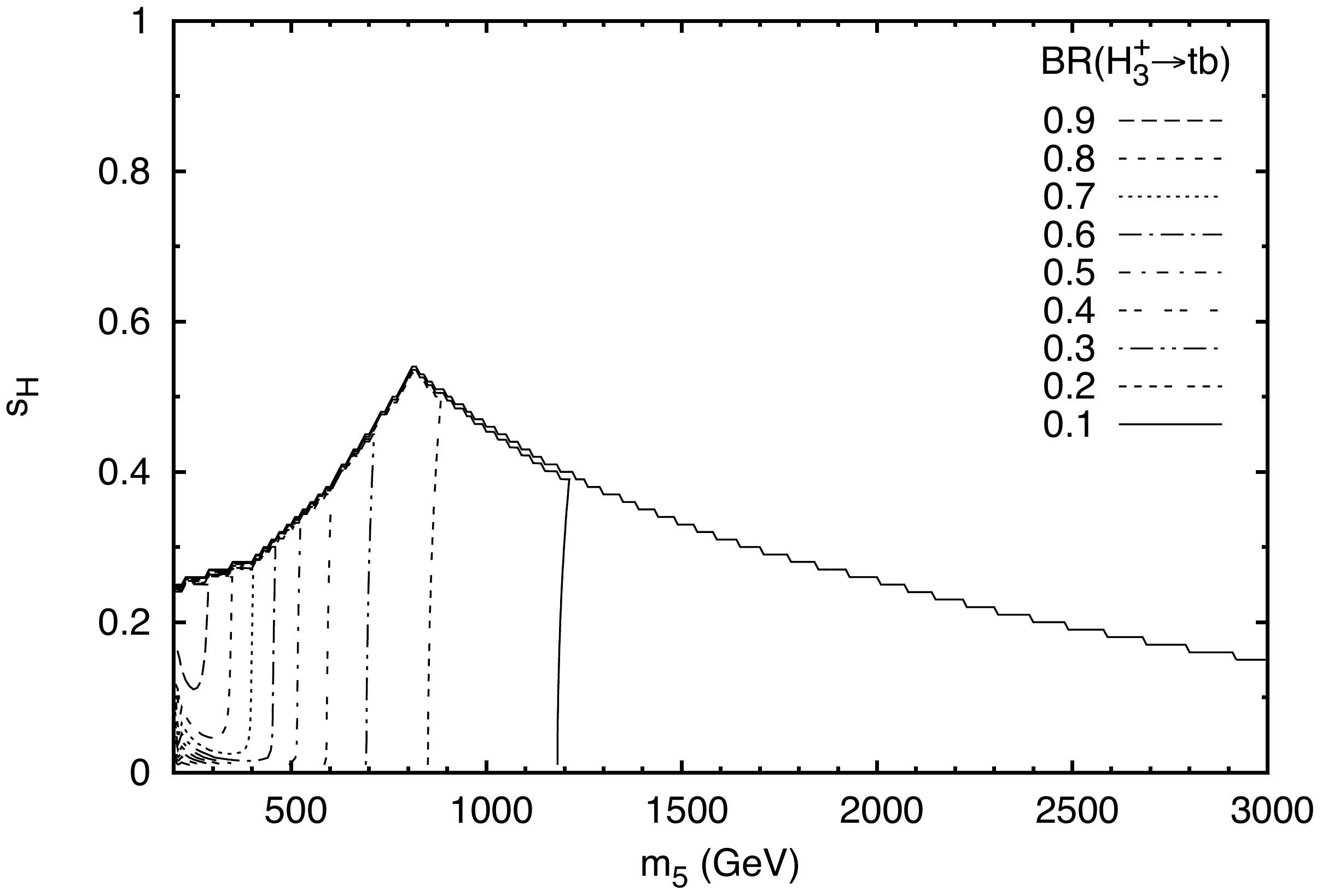}}%
\resizebox{0.5\textwidth}{!}{\includegraphics{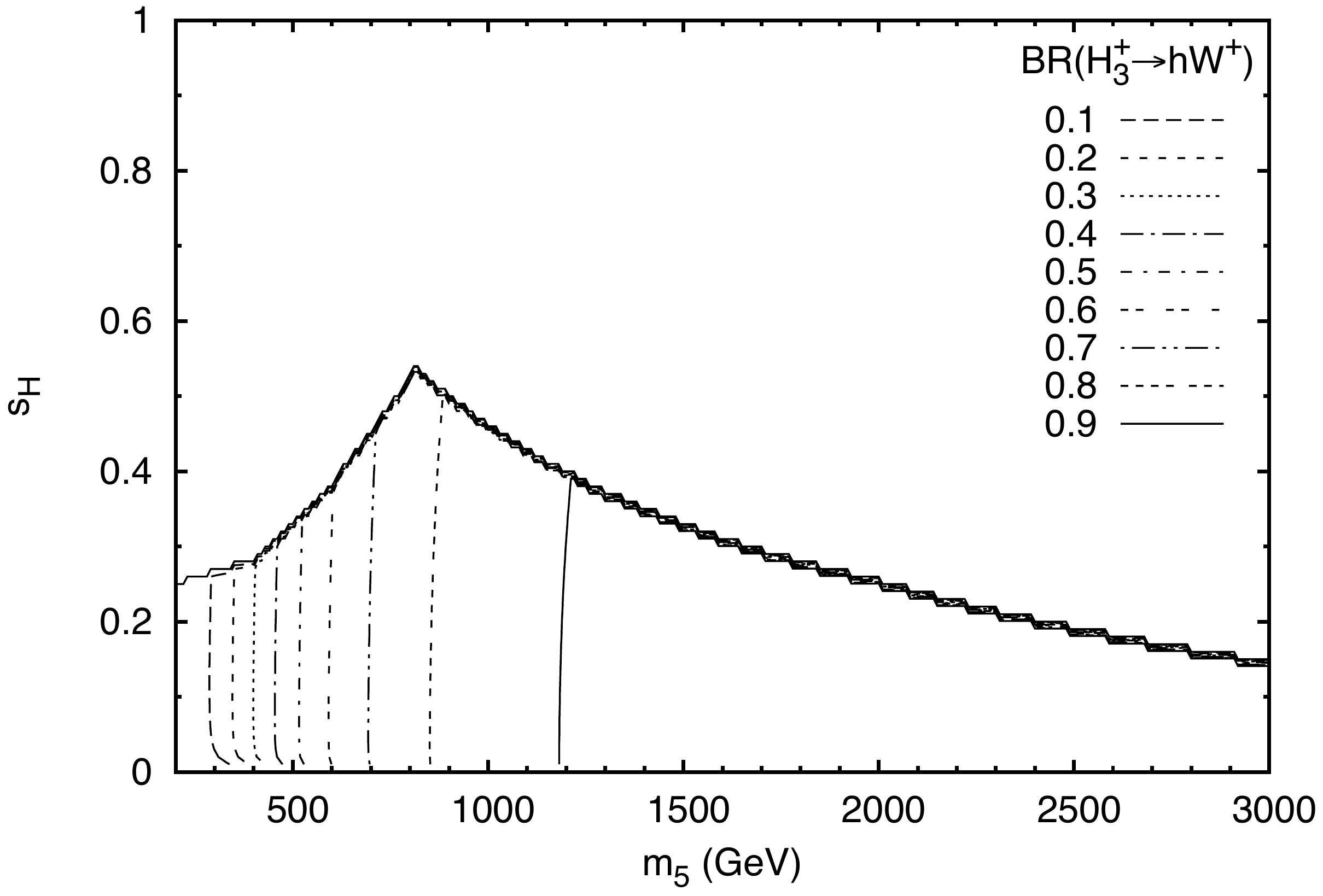}}
\caption{Contours of BR($H_3^+ \rightarrow t \bar{b}$) (left) and BR($H_3^+ \rightarrow h W^+$) (right) in the H5plane benchmark.
         BR($H_3^+ \rightarrow t \bar{b}$) ranges from $0.013$ to $0.964$ and
         BR($H_3^+ \rightarrow h W^+$) ranges from $3 \times 10^{-4}$ to $0.987$.}
\label{fig:H3PBRTB_data}
\end{figure}

\begin{figure}
\resizebox{0.5\textwidth}{!}{\includegraphics{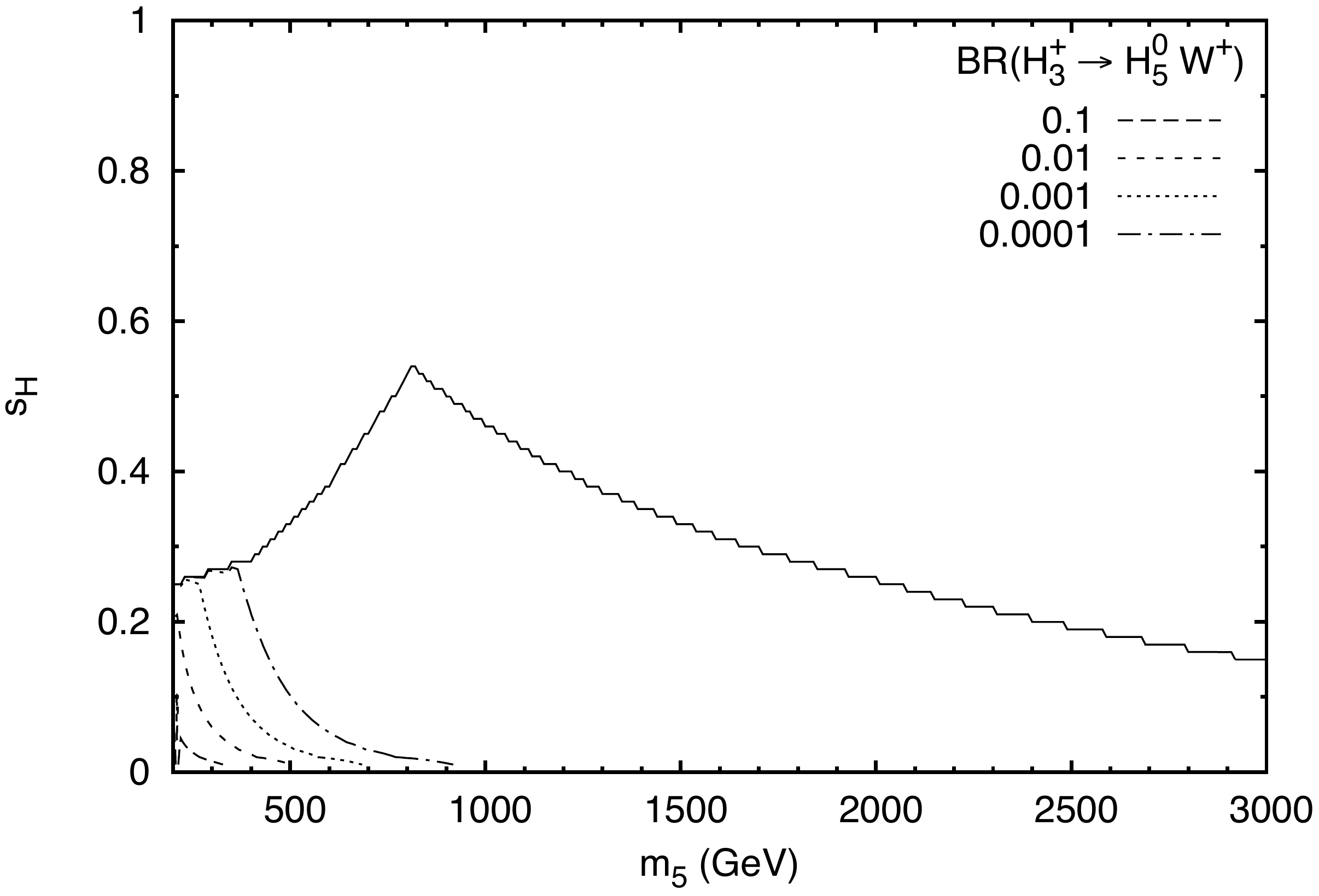}}%
\resizebox{0.5\textwidth}{!}{\includegraphics{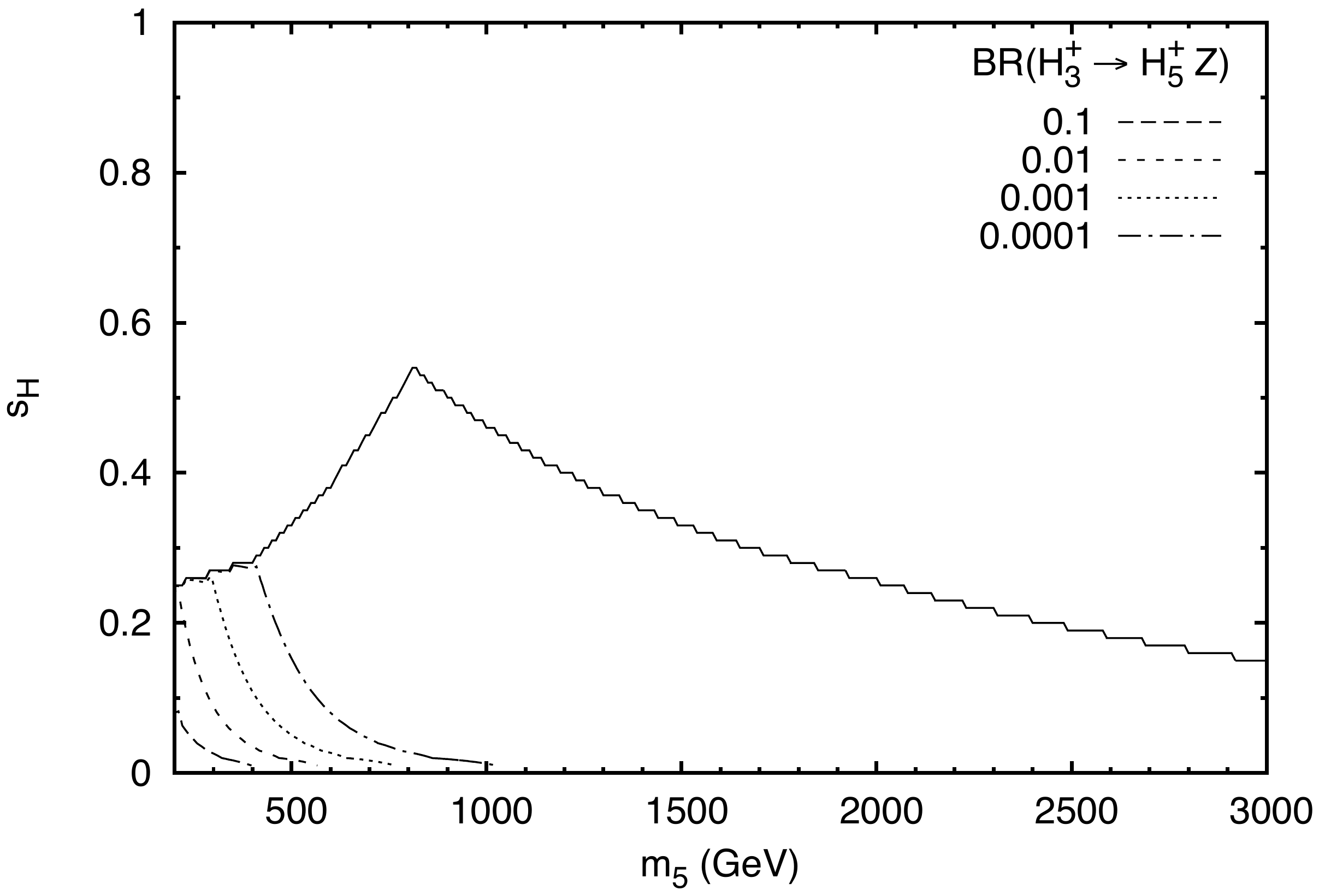}} \\
\resizebox{0.5\textwidth}{!}{\includegraphics{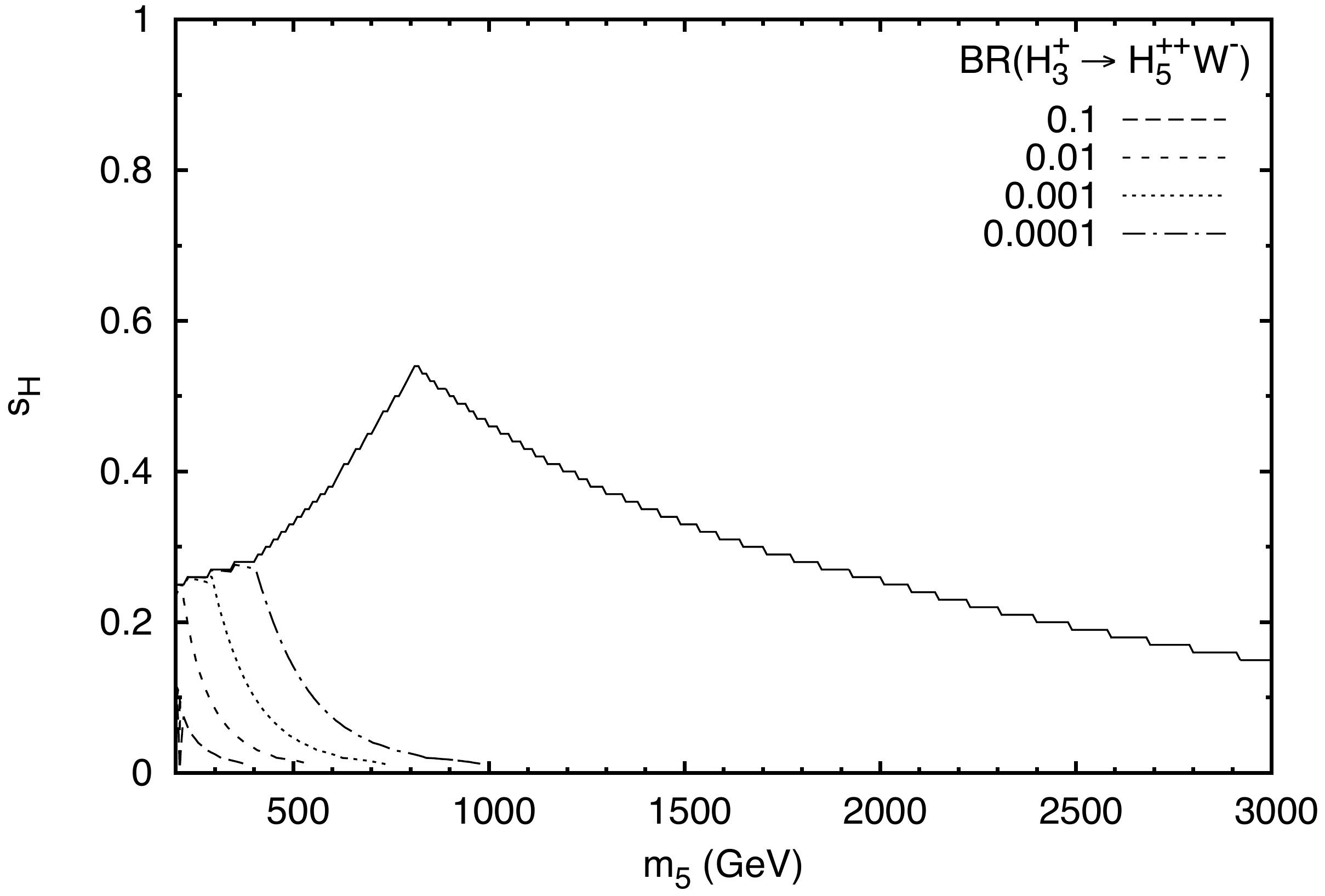}}
\caption{Contours of BR($H_3^+ \rightarrow H_5^0 W^+$) (top left), BR($H_3^+ \rightarrow H_5^+ Z$) (top right), and BR($H_3^+ \rightarrow H_5^{++} W^-$) (bottom) in the H5plane benchmark.
See text for further discussion.}
\label{fig:H3PBRWH5N_data}
\end{figure}

The dominant decays of $H_3^+$ in the H5plane benchmark are to $t \bar b$, $hW^+$, $H_5^0 W^+$, $H_5^+Z$, and $H_5^{++}W^-$.  We plot the branching ratios for these modes in Figs.~\ref{fig:H3PBRTB_data} and \ref{fig:H3PBRWH5N_data}.  The decay to $t \bar b$ dominates at low $m_5$, reaching a maximum of more than 95\% for $m_5 \sim 250$~GeV.  This branching ratio falls with increasing $m_5$ and is supplanted by the decay to $h W^+$.  The branching ratio for $H_3^+ \to h W^+$ becomes dominant ($> 50\%$) for $m_5 \gtrsim 500$~GeV and surpasses 90\% when $m_5 \gtrsim 1200$~GeV.

The branching ratios of $H_3^+$ to $H_5^0 W^+$, $H_5^+ Z$, and $H_5^{++} W^-$ are significant only for very low values of both $m_5$ and $s_H$ within the H5plane benchmark.  In this corner of parameter space, the branching ratios of these modes can be significant, reaching maxima of 25\%, 79\%, and 49\%, respectively, in slightly different regions of parameter space.  Again, though, these numbers should be treated with caution because the decays of $H_3^+$ to $H_5V$ face the same issues with the transition from on shell to off shell as the decays of $H_3^0$ to $H_5V$.  All three of these branching ratios quickly fall below the 1\% level for $m_5 \gtrsim 500$~GeV.  These decay modes also decline quickly with increasing $s_H$, due to an increase in the partial width for $H_3^+ \to t \bar b$ with increasing $s_H$.

\begin{figure}
\resizebox{0.5\textwidth}{!}{\includegraphics{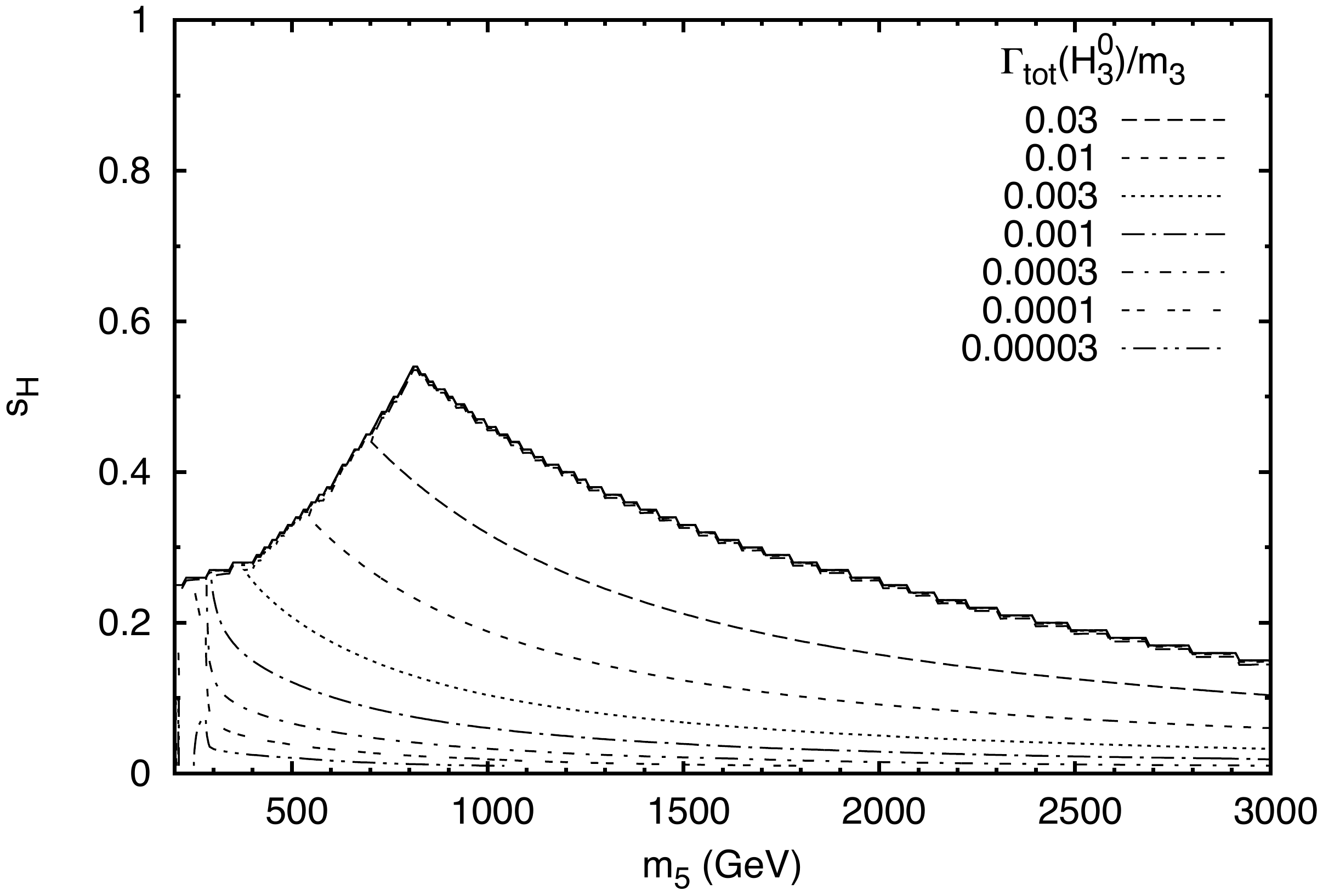}}%
\resizebox{0.5\textwidth}{!}{\includegraphics{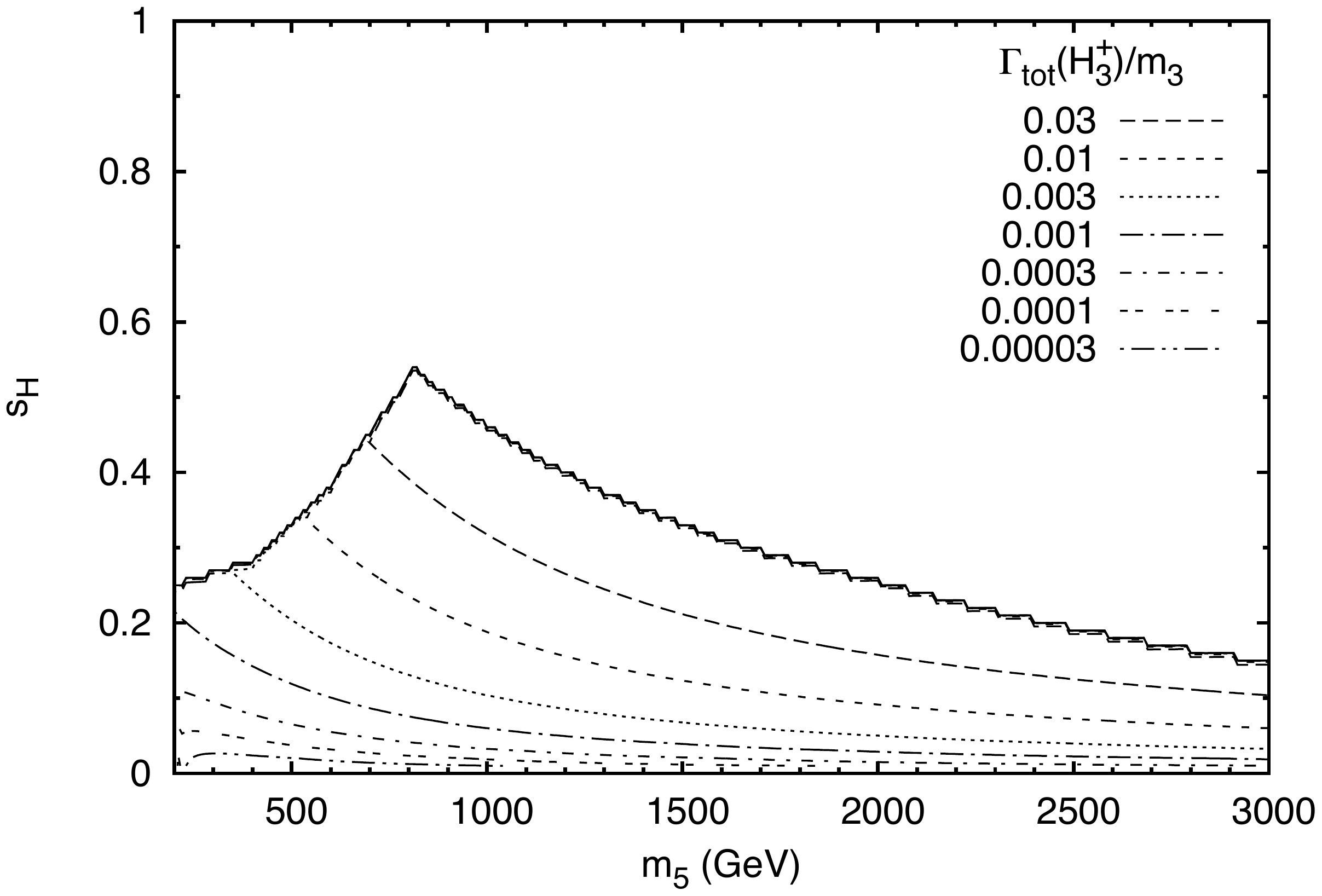}}
\caption{Contours of $\Gamma_\text{tot}(H^0_3)/m_3$ (left) and $\Gamma_\text{tot}(H^+_3)/m_3$ (right) in the H5plane benchmark.
         $\Gamma_\text{tot}(H^0_3)/m_3$ ranges from $6.6 \times 10^{-6}$ to $0.077$ and
         $\Gamma_\text{tot}(H^+_3)/m_3$ ranges from $6.2 \times 10^{-6}$ to $0.077$.}\label{fig:H3NWDTH_data}
\end{figure}

Finally, we plot the total widths of $H_3^0$ and $H_3^+$ in Fig.~\ref{fig:H3NWDTH_data}.  
They both remain quite small over the entire allowable region: although they do increase with increasing $s_H$
and $m_5$, the width-to-mass ratio $\Gamma_\text{tot}(H_3)/m_3$ never rises above 8\% for either $H_3^0$ or $H_3^+$.

\section{Conclusions}
\label{sec:conclusions}

In this paper we studied the constraints on and phenomenology of the H5plane benchmark scenario in the Georgi-Machacek model.  The H5plane benchmark has two free parameters, $m_5$ and $s_H$, where $s_H^2$ is equal to the fraction of $M_W^2$ and $M_Z^2$ that is generated by the vev of the isospin triplets.  The H5plane benchmark is defined for $m_5 \in [200,3000]$~GeV.  Existing theoretical and experimental constraints limit $s_H$ to be below 0.55 in the H5plane benchmark, so that at most 30\% of the $W$ and $Z$ boson squared-masses can be generated by the triplets.  A full parameter scan of the GM model yields an allowed region in the $m_5$--$s_H$ plane only slightly larger than in the H5plane benchmark for $m_5 \in [200,3000]$~GeV.  Our numerical work has been done using the public code {\tt GMCALC 1.2.1}.

We showed that the couplings of the 125~GeV Higgs boson $h$ in the H5plane benchmark are sufficiently SM-like that the benchmark is not further constrained by the ATLAS and CMS measurements of Higgs production and decay at LHC center-of-mass energies of 7 and 8~TeV---in fact, over most of the H5plane benchmark, the fit to LHC data is slightly better than in the SM.  Over the H5plane benchmark, compared to their SM values, the $h$ coupling to fermions can be suppressed by up to 10\% or enhanced by up to 1.4\%, its coupling to vector boson pairs can be enhanced by up to 21\%, and its loop-induced coupling to photon pairs can be suppressed by up to 1.3\% or enhanced by up to 24\% (loops involving the charged scalars in the GM model contribute non-negligibly to this).  The total width of $h$ can be suppressed by up to 2.9\% or enhanced by up to 3.5\% compared to that of the SM Higgs boson; the smallness of this range is due to an accidental cancellation among the fermionic and bosonic contributions.

By design, the mass-degenerate $H_5^{\pm\pm}$, $H_5^{\pm}$, and $H_5^0$ scalars are the lightest new scalars in the H5plane benchmark, and hence decay only to vector boson pairs at tree level.  Due to the parameter specifications in the benchmark, the mass splittings $m_3-m_5$ and $m_H-m_5$ are almost constant with $s_H$, depending primarily on $m_5$.  They fall from maxima of 84 and 120 GeV, respectively, at $m_5 = 200$~GeV to minima of 7 and 9~GeV, respectively, at $m_5 = 3000$~GeV.  (These mass splittings vary much more freely in the full GM model.)  While the mass-to-width ratios of all the new scalars in the GM model remain below 8\% in the H5plane benchmark, the fairly small mass splitting between $H_5^0$ and $H$ means that these two resonances can overlap and interfere when produced in vector boson fusion and decaying to $W^+W^-$ or $ZZ$.  Their mass splitting becomes smaller than their intrinsic widths when $m_5 \gtrsim 700$~GeV, unless $s_H$ is small.

Finally we studied the production and decays of the new heavy Higgs bosons in the GM model in the H5plane benchmark.
We found that, due to coupling suppressions, the production cross section of $H$ in gluon fusion (vector boson fusion) can be at most 58\% (4.8\%) as large as that of a SM Higgs boson of the same mass.  $H$ decays mainly to $W^+W^-$ and $ZZ$ for $m_5$ below 600--1000~GeV (depending on $s_H$), and mainly to $hh$ for $m_5$ above 700--1300~GeV.  Its branching ratio to $t \bar t$ can top 30\% for $m_5$ between 400 and 700~GeV.

$H_3^0$ decays predominantly to $t \bar t$ from the kinematic threshold at $m_5 = 280$~GeV up to $m_5 \simeq 500$~GeV, where $hZ$ takes over as the dominant decay mode.  Below the $t \bar t$ threshold, decays to $H_5^0 Z$ and $H_5^{\pm} W^{\mp}$ can be significant, but improvements to the handling of near-threshold decays in {\tt GMCALC} are needed to fully explore the branching ratios in this region.  $H_3^+$ decays predominantly to $t \bar b$ for $m_5$ values up to about 500~GeV, where $h W^+$ takes over as the dominant decay mode.


\begin{acknowledgments}
We thank Dag Gillberg for helpful conversations.
This work was supported by the Natural Sciences and Engineering Research Council of Canada.  H.E.L.\ was also partially supported through the grant H2020-MSCA-RISE-2014 no.\ 645722 (NonMinimalHiggs).
\end{acknowledgments}


\end{document}